\documentclass[aps,twocolumn,showpacs,preprintnumbers,amsmath,amssymb,nofootinbib,superscriptaddress,showkeys]{revtex4}

\usepackage{epsfig}
\usepackage{graphicx}

\begin{document}

\title{Renormalization of chiral two-pion exchange NN interactions. \\ 
  Momentum vs.\ coordinate space.} 
\author{D. R. Entem}\email{entem@usal.es}
  \affiliation{Grupo de F\'{\i}sica Nuclear, IUFFyM, Universidad de
  Salamanca, E-37008 Salamanca,Spain} 
\author{E. Ruiz
  Arriola}\email{earriola@ugr.es} \affiliation{Departamento de
  F\'{\i}sica At\'omica, Molecular y Nuclear, Universidad de Granada,
  E-18071 Granada, Spain.}
\author{M. Pav\'on
  Valderrama}\email{m.pavon.valderrama@fz-juelich.de}
  \affiliation{Institut f\"ur Kernphysik, Forschungszentrum J\"ulich,
  52425 J\"ulich, Germany} 
\author{R. Machleidt}
\affiliation{Department of Physics, University of Idaho, Moscow, Idaho 83844}
 \email{machleid@uidaho.edu} 
  \date{\today}

\begin{abstract} 
\rule{0ex}{3ex} The renormalization of the chiral np interaction in
the $^1S_0$ channel to N3LO in Weinberg counting for the long
distance potential with one single momentum and energy independent
counterterm is carried out. This renormalization scheme yields finite
and unique results and is free of short distance off-shell
ambiguities. We observe good convergence in the entire
elastic range below pion production threshold and find that there
are some small physical effects missing in the purely pionic chiral NN
potential with or without inclusion of explicit $\Delta$ degrees of
freedom. We also study the renormalizability of the standard Weinberg
counting at NLO and N2LO when a momentum dependent polynomial counterterm is
included. Our numerical results suggest that the inclusion of this
counterterm does not yield a convergent amplitude (at NLO and N2LO).

\end{abstract}


\maketitle



\section{Introduction} 

The modern Effective Field Theory (EFT) analysis of the NN interaction
using chiral symmetry as a constraint has a recent but prolific
history~\cite{Weinberg:1990rz,Ordonez:1995rz} (for comprehensive
reviews see
e.g. Ref.~\cite{Bedaque:2002mn,Epelbaum:2005pn,Machleidt:2005uz}). Most
theoretical setups are invariably based on a perturbative
determination of the chiral
potential~\cite{Rijken:1995pu,Kaiser:1997mw,Kaiser:1998wa,Friar:1999sj,Kaiser:1999ff,Kaiser:1999jg,
Kaiser:2001at,Kaiser:2001pc,Kaiser:2001dm} and the subsequent solution
of the scattering
problem~\cite{Epelbaum:1998ka,Epelbaum:1999dj,Rentmeester:1999vw,Entem:2001cg,
Entem:2002sf,Rentmeester:2003mf,Epelbaum:2003gr,Epelbaum:2003xx,Entem:2003cs,Entem:2003ft,Epelbaum:2004fk}. Actually,
the theory encounters many problems in the low partial waves and in
particular in the $s-$waves (see however Ref.~\cite{Machleidt:2006kx}
for a more optimistic view). Indeed, there is at present an ongoing
debate on how an EFT program should be sensibly implemented within the
NN context and so far no consensus has been achieved (see
e.g. \cite{Nogga:2005hy,RuizArriola:2006hc,Epelbaum:2006pt,Rho:2006tx}). The
discussion is concerned with the issue of renormalization vs.\ finite
cutoffs, a priori (power counting) vs.\ a posteriori error estimates
or the applicability of perturbation theory both on a purely short
distance theory or around some non-perturbative distorted waves. At
the moment, it seems fairly clear that an EFT scheme with a
cutoff-independent and systematic perturbative power counting for the
S-matrix (the so-called KSW counting)
fails~\cite{Kaplan:1998tg,Kaplan:1998we,Fleming:1999ee,Fleming:1999bs}.
On the other hand, the original EFT inspired
scheme~\cite{Weinberg:1990rz,Ordonez:1995rz} (the so-called Weinberg
counting) has recently been shown to produce many results which turn
out to be strongly cutoff
dependent~\cite{Nogga:2005hy,Valderrama:2005wv} and hence to be
incompatible with renormalizability~\footnote{In this paper we refer
to renormalizability in the sense of parameterizing the short distance
physics by a potential which matrix elements in momentum space are a
polynomial in the momenta.}. Thus, some acceptable compromise must be
made. Actually, the chosen approach between this dichotomy depends
strongly on the pursued goals and it is fair to say that any choice
has both advantages and disadvantages.  In any case, the reason for
both failures can be traced back to the nature of the long-distance
chiral potentials; while pion-exchange potentials fall off
exponentially at long distances they include strong power-law
singularities at short distances. Those singularities become
significant already at distances comparable with the smallest de
Broglie wavelength probed in NN scattering below pion production
threshold. Obviously, any development of NN interactions based on
chiral dynamics will presumably require a deeper understanding and
proper interpretation of the peculiarities of these highly singular
chiral potentials. Although singular potentials where first analyzed
many years ago~\cite{Case:1950an} (for an early review see
e.g.~\cite{Frank:1971xx} and for a more updated view within an EFT
context see~\cite{Beane:2000wh}), their short distance singular
character within the NN interaction has seriously been faced more
recently within a renormalization context for the one-pion exchange
(OPE) potential~\cite{PhysRevC.49.2942,Frederico:1999ps,
Beane:2001bc,Eiras:2001hu,PavonValderrama:2004nb,
PavonValderrama:2005gu,Nogga:2005hy,
Birse:2005um,Epelbaum:2006pt,Yang:2007hb} and the two-pion exchange
(TPE)
potential~\cite{PavonValderrama:2004td,Valderrama:2005wv,PavonValderrama:2005uj}.

In the np scattering problem the $^1S_0$ channel is very special since
the scattering length is unnaturally large as compared to the range of
the strong interaction, $\alpha_0 = -23.74(2)\: {\rm fm} \gg 1/m_\pi =1.4
\: {\rm fm}$. In fact, even at zero energy the wave function probes
relatively short distance components of the chiral
potential~\cite{Valderrama:2005wv}~\footnote{This can be best seen by
means of the effective range formula
$$ r_0 = 2 \int_0^\infty dr \left[ u_0 (r)^2 - \left( 1-
\frac{r}{\alpha_0} \right)^2 \right] 
$$ where $u_0 (r)$ is the zero energy wave function, fulfilling the
asymptotic condition $u_0 (r) \to 1 - r/\alpha_0$. Most of the
integrand is located in the region around $r=1 {\rm fm}$ which is in
between OPE and TPE ranges. Moreover, the low energy theorem of
Ref.~\cite{Valderrama:2005wv} allows to write $r_0 = A + B/\alpha_0 +
C/\alpha_0^2$ which in the extreme limit $\alpha_0 \to \infty$ yields
$r_0 \to A$.  Numerically it is found that $A$ is far more dependent
than $B$ and $C$ when being evaluated at LO, NLO and NNLO.}. Higher
energies become even more sensitive to short distance interactions.
Consequently, this channel looks like an ideal place to learn about
the size of the most relevant short range corrections to the NN force
in the elastic scattering region. Actually, in the $^1S_0$ channel,
most EFT inspired schemes yield at leading order (LO) (which consists
of OPE plus a nonderivative counterterm) an almost constant phase of
about $75^o$ around $k=250 \: {\rm MeV}$, and an effective range of
$r_0^{\rm LO} = 1.44 {\rm fm}$. On the other hand, most determinations
from Partial Wave Analysis~\cite{Stoks:1993tb,Arndt:1994br} and high
quality potential models~\cite{Stoks:1994wp} yield an almost vanishing
phase at this center-of-mass (CM) momentum while the experimental
effective range is about twice the OPE value, $r_0^{\rm exp} = 2.77
(5) {\rm fm}$.~\footnote{In fact, these high quality potential models
yield slightly smaller values, $r_0 \approx 2.67 {\rm fm}$.}  It is
quite unbelievable that such large changes can be reliably
accommodated by perturbation theory starting from this LO result
despite previous unsuccessful attempts treating OPE and TPE
perturbatively~\cite{Kaplan:1998tg,Kaplan:1998we,Fleming:1999ee,
Fleming:1999bs}.
Actually, for the singlet channel case, 
short distance components are enhanced due to the large value of the
scattering length and the weakness of the OPE interaction in this channel.  
This is why TPE contributions have been treated
with more success in a non-perturbative
fashion~\cite{Ordonez:1995rz,Epelbaum:1998ka,Epelbaum:1999dj,Rentmeester:1999vw,Entem:2001cg,
Entem:2002sf,Rentmeester:2003mf,Epelbaum:2003gr,Epelbaum:2003xx,Entem:2003cs,Entem:2003ft,Epelbaum:2004fk}. Despite
phenomenological agreement with the data, the inclusion of finite
cutoffs suggests that there might be some regulator dependence in
those calculations.

In a series of recent
papers~\cite{PavonValderrama:2005gu,Valderrama:2005wv,PavonValderrama:2005uj},
two of us (M.P.V. and E.R.A.) have proposed not only to iterate but
also to renormalize to all orders the NN chiral potential within a
long distance expansion. By allowing the minimal number of
counterterms to yield a finite result, long-distance regulator-independent
correlations are established.  In practice, the potential must be
computed within some power counting scheme. While the potential is
used within Weinberg's power counting to LO, next-to-leading order
(NLO), and next-to-next-to-leading order (N2LO), we only allow for
those counterterms which yield a finite and unique scattering
amplitude. In the $^1S_0$ np channel a single energy and momentum
independent counterterm $C_0$ is considered which is determined by
adjusting the physical scattering length. In the derivation of this
result the mathematical requirements of completeness and
self-adjointness for the renormalized quantum mechanical problem for a
local chiral potential play a decisive role. This surprising result is
in contrast to the standard Weinberg counting where an additional
counterterm $C_2$ is included already at NLO. This $C_2$ counterterm
could, in fact, be determined by fitting the experimental value of the
effective range; the physics of $C_2$ is to provide a short distance
contribution to the effective range in addition to the contribution
from the known long distance chiral potential. Within this context it
is remarkable that according to Ref.~\cite{Valderrama:2005wv}, where
such a short distance contribution vanishes (or equivalently $C_2=0$
when the cut-off is removed), rather accurate values are {\it
predicted} yielding $r_0^{\rm NLO}=2.29\: {\rm fm}$ and $r_0^{\rm
N2LO}=2.86\: {\rm fm}$ {\it after renormalization}. This latter value
is less than $3\%$ larger than the experimentally accepted value and
it suggests that most of the effective range is saturated by N2LO TPE
contributions and calls for pinning down the remaining
discrepancy. This trend to convergence and agreement is also shared by
higher order slope parameters in the effective range expansion without
strong need of specific counterterms although there is still room for
improvement.  Motivated by this encouraging result, one goal of the
present paper is to analyze the size of the
next-to-next-to-next-to-leading order (N3LO) corrections to the
results found in~\cite{Valderrama:2005wv}.

The calculations in
Ref.~\cite{PavonValderrama:2005gu,Valderrama:2005wv,PavonValderrama:2005uj,Valderrama:2007nu}
exploit explicitly the local character of the chiral potential by
conducting the calculations in coordinate space 
which makes the analysis more transparent.
Many results, in particular the conditions under which a renormalized
limit exists, can be established a priori analytically. Moreover, the
highly oscillatory character of wave functions 
at short distances 
is treated numerically using efficient adaptive-step
differential equations techniques. This situation contrasts with
momentum space calculations where, with the exception of the pion-less
theory, there is a paucity of analytical results, and one must
mostly rely on numerical methods.
Moreover, the existence of a
renormalized limit is not obvious a priori and one may have to resort to
some trial and error to search for counterterms. 
Finally, renormalization conditions
are most naturally formulated at zero energy for which the momentum
space treatment may be challenging, at times. 
Of course, besides these technical issues,
there is no fundamental difference between
proceeding in momentum or coordinate space, particularly {\it after}
renormalization, provided the same renormalization conditions are
specified, since disparate regulators stemming from either space are
effectively removed.  Indeed, we will check agreement for the
phase-shifts determined in different spaces whenever such a comparison
becomes possible. This equivalence is in itself a good motivation for
renormalization.

However, at N3LO some unavoidable non-localities appear in the chiral
long distance potential. Although they could be treated in
configuration space, we adopt here a momentum space treatment. This
will also allow us to answer an intriguing question which was left
open in the coordinate space analysis of previous
works~\cite{PavonValderrama:2005gu,Valderrama:2005wv,PavonValderrama:2005uj},
namely, the role played by the conventional momentum-polynomial
representations of the short distance interaction used in most
calculations~\cite{Epelbaum:1998ka,Epelbaum:1999dj,Entem:2001cg,
Entem:2002sf,Rentmeester:2003mf,Epelbaum:2003gr,Epelbaum:2003xx,Entem:2003cs,Entem:2003ft,Epelbaum:2004fk}
in the renormalization problem. More specifically,
Ref.~\cite{Valderrama:2005wv} showed that taking $C_2=0$ was
consistent, and a regularization scheme exists where a fixed $C_2$ was
irrelevant, but could not discriminate whether $C_2 \neq 0 $ was
inconsistent as far as it was readjusted to the effective range
parameter for any cut-off value.~\footnote{We mean of course the case
when the cut-off is being removed. The essential issue is whether or
not one can fix by a short distance potential which is a polynomial in
the momenta the effective range {\it independently} on the potential
and remove the cut-off at the same time. Of course, the very
definition of the potential is ambiguous and requires a specific
choice on the polynomial parts. Technically, we find that any {\it
fixed}, cut-off indendependent $C_2$, becomes irrelevant in the limit
$\Lambda \to \infty$ (see below).} In this regard, the present paper
yields a definite answer making the surprising agreement of the
effective range found in Ref.~\cite{Valderrama:2005wv} an inevitable
mathematical consequence of renormalization.

The paper is organized as follows. The problem is stated in
Sec.~\ref{sec:renorm-prob} where a general overview of the
renormalization problem is given both in momentum as well as
coordinate space. In Sec.~\ref{sec:mom-space}, we particularize the
momentum space formulation of the scattering problem with counterterms
for the $^1S_0$ channel within a sharp three-momentum cutoff scheme.
Likewise, in Sec.~\ref{sec:coor-space} we proceed similarly in the
coordinate space formulation within a boundary condition
regularization with a short distance cutoff $r_c$.  In
Sec.~\ref{sec:coor-mom}, we discuss some features of both coordinate
and momentum space formulations in the pion-less theory and try to
connect the high momentum cutoff $\Lambda$ with the short distance
radial cutoff $r_c$. This allows a one-to-one mapping of counterterms
in both spaces which will prove useful later on in the pion-full
theory. The identification between the sharp momentum cutoff and the
short distance radius found in the pion-less theory is discussed
further in Appendix~\ref{sec:nyquist} in the presence of a long
distance potential in the light of the Nyquist theorem.  In
Sec.~\ref{sec:N3LO}, we come to the central discussion on N3LO
corrections to the phase shifts when the scattering amplitude is
renormalized with only one short distance counterterm. A wider
perspective is achieved by further considering the role of explicit
$\Delta$-excitations in intermediate states and the subsequent
one-counterterm renormalization of the scattering amplitude. We also
discuss the role of three pion exchange as well as how the results
depend on the renormalization scheme used to compute the potential
based on cut-off independent counterterms. The mathematical
justification for using just one counterterm is provided in
Sec.~\ref{sec:weinberg} where the standard Weinberg scheme is pursued
both in momentum and coordinate space at NLO and N2LO and shown to
potentially have some problems.  Finally, in Sec.~\ref{sec:conc}, we
summarize our main points.

\section{The renormalization problem}
\label{sec:renorm-prob}

\subsection{General overview and main results} 
\label{sec:overview} 

Let us define the scope and goals of the present work.  The standard
non-perturbative formulation of the renormalization problem starts
with an effective Lagrangian or Hamiltonian~(see
e.g. \cite{Weinberg:1990rz,Ordonez:1995rz} and
\cite{Bedaque:2002mn,Epelbaum:2005pn} and references therein), 
from which a certain set of irreducible Feynman diagrams
(usually up to a certain order) is calculated. These irreducible
diagrams are defined to represent a potential $V$.
The potential is then inserted into a scattering equation
where it is iterated infinitely many times or, in other words,
re-summed non-perturbatively.
In the CM
frame, where the np kinetic energy is given by $E = p^2 / M $, with $M
= 2 \mu_{np} = 2 M_n M_p /(M_p + M_n) $, the scattering process is
governed by the Lippmann-Schwinger equation
\begin{eqnarray}
T = V + V G_0 T  \, , 
\end{eqnarray}
with $V$ the potential operator and $G_0 = (E-H_0 )^{-1} $ the
resolvent of the free Hamiltonian. The outgoing boundary condition
corresponds to  $ E \to E + {\rm i} 0^+ $. Using the
normalization $ \langle \vec x | \vec k \rangle = e^{{\rm i} \vec k
\cdot \vec x} /(2\pi)^{3/2} $ one has
\begin{eqnarray}
\langle \vec k' | T(E) | \vec k \rangle = \langle \vec k' | V | \vec k
\rangle + \int^\Lambda d^3 q { \langle \vec k' | V | \vec q \rangle
\langle \vec q | T(E) | \vec k \rangle \over E - (q^2 / 2 \mu ) } \, . 
\nonumber \\
\end{eqnarray} 
Here $\Lambda$ means a generic regulator and represents the scale
below which all physical effects are taken into account {\it
explicitly}. The degrees of freedom which are above $\Lambda$ are
taken into account {\it implicitly} by including a suitable cutoff
dependence in the potential. The precise equation governing this
cutoff dependence was described and studied in some detail in
Ref.~\cite{Birse:1998dk} with particular emphasis on infrared fixed
points. We will analyze the cutoff dependence below focusing
on the ultraviolet aspects of the interaction.

Motivated by the low energy nature of the effective theory, the
potential is usually separated into short and long distance components
in an {\it additive} form
\begin{eqnarray} 
\langle \vec k' | V | \vec k \rangle = V_S (\vec k', \vec k) + V_L ( \vec k' , 
\vec k) \, , 
\label{eq:long-short} 
\end{eqnarray} 
where the long distance contribution is usually given by successive
pion exchanges
\begin{eqnarray} 
V_L ( \vec k' , \vec k) = V_{1\pi} ( \vec k' , \vec k) + V_{2\pi} ( \vec k' ,
\vec k) + \dots \, , 
\label{eq:long} 
\end{eqnarray} 
and the short distance component is characterized by a power series
expansion in momentum
\begin{eqnarray} 
V_S (k',k) = C_0 + C_1 \vec k \cdot \vec k' + C_2 (\vec k^2 + \vec
{k'}^2 ) + \dots \,\,  ,  
\label{eq:short}  
\end{eqnarray} 
where for simplicity we assume a spin singlet channel~\footnote{More
detailed expressions including spin triplet channels can be looked up
e.g.\ in Ref.~\cite{Epelbaum:2005pn}.}. Note that $C_0$ and $C_2$
contribute to s-waves while $C_1$ contributes to p-waves, and so
on. One should face the fact that, although the decomposition given by
Eq.~(\ref{eq:long-short}) is perturbatively motivated and seems quite
natural, the additivity between short and long range forces is
actually an {\it assumption} which has important consequences, as we
will see.~\footnote{Specifically, Eq.~(\ref{eq:long-short}) does not
foresee for instance terms of the form $ V (p',p) C(\Lambda) $, i.e.
terms which are not polynomial but influence the renormalization
process. Of course, once additivity is relaxed there are many
possible representations, in particular for the short distance
components.}

In principle, for a given regularization scheme characterized by a
cutoff $\Lambda$, the counterterms $C_0$, $C_1$, $C_2$ and so on are
determined by fixing some observables. One naturally expects that the
number of renormalization conditions coincides with the number of
counterterms in a way that all renormalization conditions are fully
uncorrelated. The statement of UV-renormalizability is that such a
procedure becomes always possible when the cutoff $\Lambda$ is
removed by taking the limit $\Lambda \to \infty$. This may not be the case
as there may appear redundant contributions (see the discussion below
in Sec.~\ref{sec:coor-mom}) meaning that one counterterm or
counterterm combination can take any value. Another possible situation
is just the opposite; one may want to impose more renormalization
conditions than possible. In this case some counterterms or
counterterm combinations are forbidden. Rather than being intricate
mathematical pastimes, these features have already been investigated
recently~\cite{Nogga:2005hy,Valderrama:2005wv} for large cutoffs
casting some doubt on the
regulator independence of the original
proposal~\cite{Weinberg:1990rz,Ordonez:1995rz}.

Even if one admits generically Eq.~(\ref{eq:long-short}) as well as
Eq.~(\ref{eq:long}) and Eq.~(\ref{eq:short}), it is not obvious how
many terms should be considered and whether there is a clear way of
defining a convergence criterium or identifying a convergence
pattern. It is fairly clear that, on physical grounds, one should
consider an expansion of the potential that starts at
long distance and decreases in range 
as the number of exchanged particles increases.
Note, however, that, while OPE is well-defined, 
the general form of the TPE potential is not uniquely determined. 
Within this
context, one of the main attractive features of the EFT approach has
been the definition of a power counting scheme which provides a hierarchy and
an {\it a priori} correlation between long and short range physics. In
the present paper, we will assume Weinberg power
counting~\cite{Rijken:1995pu,Kaiser:1997mw,Kaiser:1998wa,Friar:1999sj,Kaiser:1999ff,Kaiser:1999jg,
Kaiser:2001at,Kaiser:2001pc,Kaiser:2001dm} (see below for a more
precise definition), where the long distance potential is determined in
a dimensional power expansion.

The long distance component $V_L$ is obtained by particle exchanges
and, in some simple cases, depends only on the momentum
transfer~\footnote{This assumption will be relaxed immediately below
when discussing the Weinberg counting in the $^1S_0$ channel.}.  In
such a case, if we formally take a Fourier transformation of the long
distance potential
\begin{eqnarray}
V_L (\vec x) = \int \frac{d^3 p}{(2\pi)^3} V_L (\vec p) e^{i \vec p
\cdot \vec x} \, , 
\end{eqnarray} 
and take the limit $\Lambda \to \infty $ one has the standard
Schr\"odinger equation in coordinate space,
\begin{eqnarray}
-\frac1 M \nabla^2 \Psi_k ( \vec x) + V (\vec x) \Psi_k (\vec x) = E
 \Psi_k (\vec x) \, , 
\end{eqnarray} 
where the coordinate space potential is 
\begin{eqnarray}
V (\vec x) &=& V_L (\vec x) + C_0 \delta^{(3)} (\vec x) + C_1 \vec \nabla \delta^{(3)} (\vec x) \vec \nabla
\nonumber \\ &+&
C_2 \left[ \nabla^2 \delta^{(3)} (\vec x) + \delta^{(3)} (\vec x)
\nabla^2 \right] +  
\dots \, \, . 
\label{eq:pot-coor}
\end{eqnarray} 
The whole discussion, which has been carried on for years now, concerns the
precise meaning of these delta and derivatives of delta interactions,
particularly when a long distance potential is added to the short
distance one. A crucial finding of the present paper based on a direct
analysis in momentum space is that {\it non-perturbative
renormalization imposes restrictions} on the number of terms and form of the
short distance potential which depend also on the particular long
distance potential. Some of these restrictions were discussed in
previous
works~\cite{PavonValderrama:2005gu,Nogga:2005hy,Valderrama:2005wv}. Remarkably
these new renormalizability restrictions were conjectured in
coordinate space in
Ref.~\cite{PavonValderrama:2005gu,Valderrama:2005wv,RuizArriola:2006hc}
for the $^1S_0$ channel based on self-adjointness and completeness of
states and apply to the TPE potential; a single $C_0 \neq 0$
counterterm is allowed while two counterterms, $C_0 \neq 0$ and $C_2 \neq 0$,
are forbidden~\footnote{Again, we mean cut-off dependent counterterms designed to fit physical observables.}.

\subsection{Momentum space formulation}
\label{sec:mom-space}

In the $^1S_0$ channel the scattering
process is governed by the Lippmann-Schwinger equation
\begin{eqnarray}
T (k',k) = V (k',k) + \int_0^\Lambda dq V (k',q)
\frac{q^2 M }{p^2-q^2+i 0^+} T (q,k) \, , \nonumber \\ 
\label{eq:LS} 
\end{eqnarray} 
where $T (k',k) $ and $V(k',k)$ are the scattering amplitude and the
potential matrix elements, respectively, between off-shell momentum
states $k$ and $k'$ in that channel and the sharp three-momentum
cutoff $\Lambda$ represents the scale below which all physical
effects are taken into account {\it explicitly}.  From the on-shell
scattering amplitude the phase shift can be readily obtained
\begin{eqnarray} 
T (p,p) =  -\frac{2}{ \pi M p }e^{i \delta} \sin \delta (p) \, . 
\end{eqnarray} 
The short range character of the nuclear force implies that at low
energies one has the effective range expansion (ERE)
\begin{eqnarray} 
p \cot \delta (p) &=& - \frac{2}{M\pi}\,{\rm Re}\left[ \frac{1}{T(p,p)}\right] 
\nonumber \\ 
&=& -\frac1{\alpha_0}
  + \frac12 r_0 p^2 + v_2 p^4 + v_3 p^6 + \dots
\label{eq:ERE}
\end{eqnarray} 
where $\alpha_0$ is the scattering length, $r_0$ the effective range
and $v_2$, $v_3$ etc. are slope parameters.

In the $^1S_0$ channel, the potential is decomposed as the sum of
short and long range pieces
\begin{eqnarray} 
V (k',k) = V_S (k', k) + V_L (k',k) \, . 
\label{eq:long-short-1S0} 
\end{eqnarray} 
In the standard Weinberg counting, the short distance contribution is
written as follows 
\begin{eqnarray} 
V_S (k',k) &=& C_0 (\Lambda) \nonumber + (k^2 + {k'}^2) C_2 (\Lambda)
\nonumber \\ &+& C_4' (\Lambda) k^2 {k'}^2 + C_4 (\Lambda) (k^4 + {k'}^4 ) + \dots
\label{eq:Vs(k,k')}
\end{eqnarray} 
where the counting is related to the order of the momentum which
appears explicitly. The long distance component of the potential is
taken to be the sum of explicit pion exchanges
\begin{eqnarray} 
V_L  = V_{1\pi} + V_{2\pi} + V_{3\pi}+ \dots 
\label{eq:potpi}
\end{eqnarray} 
where~\cite{Weinberg:1990rz}
\begin{eqnarray} 
V_{1\pi} &=& V_{1\pi}^{(0)} + V_{1\pi}^{(2)} + V_{1\pi}^{(3)}+
V_{1\pi}^{(4)}+ \dots \nonumber \\  
V_{2\pi} &=& V_{2\pi}^{(2)} + V_{2\pi}^{(3)} + V_{2\pi}^{(4)}+ \dots \nonumber \\  
V_{3 \pi} &=& V_{3\pi}^{(4)} + \dots \nonumber \\ 
\label{eq:1pi2pi3pi} 
\end{eqnarray} 
using dimensional power counting.  Ideally, one should determine the
physically relevant long range regulator independent correlations, i.e.,
long distance effects of similar range. This would amount to consider
{\it all} $n\pi$ exchange effects on the same footing, since they
yield a long distance suppression $\sim e^{-n m_\pi r}$ modulo power
corrections.  At present, the only way how these long distance
potentials can be systematically computed is by dimensional power
counting in perturbation theory, as represented schematically in
Eqs.~(\ref{eq:potpi}) and (\ref{eq:1pi2pi3pi}).

In the standard Weinberg counting one has 
\begin{eqnarray} 
V_{\rm LO} &=& V_S^{(0)} + V_{1\pi}^{(0)} \nonumber \\ V_{\rm NLO} &=& V_{\rm
LO} + V_S^{(2)} + V_{1\pi}^{(2)} + V_{2\pi}^{(2)} \nonumber \\ V_{\rm N2LO} &=&
V_{\rm NLO} + V_{1\pi}^{(3)} + V_{2\pi}^{(3)} \nonumber \\ V_{\rm N3LO} &=&
V_{\rm N2LO} + V_S^{(4)} + V_{1\pi}^{(4)} + V_{2\pi}^{(4)}  +
V_{3\pi}^{(4)} \nonumber \\ 
\end{eqnarray} 
Note that this counting involves both unknown short-distance physics
and chiral long-distance physics in a {\it uncorrelated way}. Note
also that there is no first order contribution and that there is no
third order contribution to the short distance potential.

Regarding Eq.~(\ref{eq:long-short-1S0}) one should stress that the
separation between long and short range contributions to the potential
is not unique. In fact, there is a polynomial ambiguity in the long
range part which can freely be transferred to the short distance
contribution. However, the non-polynomial part is unambiguous as it is
directly related to the left hand cut of the partial wave amplitude
which for $n \pi $ exchange is located at $p= i n m_\pi /2$ but presumably
becomes incomplete for $|p|> m_\rho/2$.~\footnote{The best way to
recognize the ambiguity is in terms of the spectral function
representation of the potential~\cite{Kaiser:1997mw}, where the
subtraction constants can be fixed arbitrarily. In coordinate space
the non-ambiguous part corresponds to the potential $V(r)$ for any
non-vanishing radius (see e.g. the discussion in
Ref.~\cite{Valderrama:2005wv,PavonValderrama:2005uj}.)}

\subsection{Coordinate space formulation}
\label{sec:coor-space} 

In coordinate space, the problem in the $^1S_0$ channel is formulated
as follows~\cite{PavonValderrama:2004td,Valderrama:2007nu}. Assuming a
{\it local} long distance potential $V_L (r)$ one has to solve the
Schr\"odinger equation
\begin{equation} -u_p '' (r) + U_L (r) u_p (r) = p^2 u_p (r)\, ,
\qquad r > r_c \, , 
\label{eq:sch_k} 
\end{equation}
where $U_L (r) = 2 \mu_{np} V_L (r) $ is the reduced potential (in
fact, the Fourier transformation of $V_L(q)$ ) and $u_p(r)$ the
reduced wave function for an s-wave state. Here $r_c$ is the short
distance cutoff and the reduced wave function is subject to the
boundary condition at $r=r_c$ and the standard long distance free
particle behaviour
\begin{eqnarray}
\frac{u_p' (r_c)}{u_p (r_c)} &=& p \cot \delta_S (p)  \, ,\\ 
u_p (r) &\to& \frac{\sin (p r + \delta(p))}{\sin \delta(p)} \, .
\label{eq:bc_sch} 
\end{eqnarray}
where $\delta_S (p)$ is the short distance phase-shift encoding
the physics for $r < r_c$. In the case of a vanishing long range
potential $U_L (r)=0$ the phase shift is given by $\delta_S(p,r_c)$. On
the other hand, if we take $\delta_S(p)=0$ we get a standard problem
with a hard core boundary condition, $u_p(r_c)=0$ which for $r_c\to 0
$ becomes the standard regular solution at the origin. At low energies
both the {\it full} phase-shift $\delta(p) $ and the {\it short
distance} phase-shift $\delta_S(p)$ can be described by some low
energy approximation, like e.g., an effective range expansion,
\begin{eqnarray}
p \cot \delta_S (p) &=& - \frac{1}{\alpha_{0,S}}+ \frac12 r_{0,S} p^2 +
\dots
\label{eq:ere_short} \\ 
p \cot \delta (p) &=& - \frac{1}{\alpha_{0}}+ \frac12 r_{0} p^2 +
\dots
\label{eq:ere_full}
\end{eqnarray} 
where $ \alpha_{0,S}$ is the short range scattering length, $ r_{0,S}$
the short range effective range, and $\alpha_0$ and $r_0$ the full
ones~\footnote{This is not the only possible short distance
representation~\cite{Valderrama:2007nu}. See
Sec.~\ref{sec:coor-pion-less} for a further discussion on this.}. If
we also make an expansion at low energies of the reduced wave function
\begin{eqnarray}
u_p (r)= u_0 (r) + p^2 u_2 (r) + \dots   
\end{eqnarray} 
we get the hierarchy of equations 
\begin{eqnarray} 
-u_0 '' (r) + U(r) u_0 (r) &=& 0  \, , \qquad \label{eq:u0} \\ 
\alpha_{0,S} u_0' (r_c) +  u_0\,(r_c) &=& 0 \, , \nonumber \\  
u_0 (r) &\to&   1- \frac{r}{\alpha_0}  \, , \nonumber
\end{eqnarray} 
and 
\begin{eqnarray} 
-u_2 '' (r) + U(r) u_2 (r) &=& u_0 (r) \, , \label{eq:u2} \\
\alpha_{0,S} u_2' (r_c) + u_2 (r_c) &=& \frac12 r_{0,S} \alpha_{0,S}
u_0 (r_c) \, , \nonumber \\ u_2 (r) &\to& \frac{r}{6 \alpha_0}\left(r^2
-3 \alpha_0 r + 3 \alpha_0 r_0 \right) \, , \nonumber
\end{eqnarray} 
and so on. The standard way to proceed would be to integrate the
equations for $u_0(r)$, $u_2(r)$, etc. from infinity downwards, with a
known value of $\alpha_0 $, using Eq.~(\ref{eq:u0}) to obtain
$\alpha_{0,S}$ and then one can use Eq.(\ref{eq:sch_k}) together with
Eq.~(\ref{eq:bc_sch}) and Eq.~(\ref{eq:ere_short}) to compute $
\delta(k) $ for any energy with a given truncated boundary
condition. This procedure provides by construction the low energy
parameters we started with and takes into account that the long range
potential determines the form of the wave function at long
distances. The only parameter in the procedure is the short distance
radius $r_c$, which is eventually removed by taking the limit $r_c \to
0$.

One should mention at this point that the coordinate space is
particularly suited for the case of local long distance potentials,
but the renormalization with an arbitrary number of countertersm
requires an {\it energy dependent} but real boundary condition at
short distances which eventually violates self-adjointness. On the
other hand, the momentum space formulation allows the discussion of
nonlocal long distance potentials and the renormalization is done in
terms of a {\it momentum dependent} short distance polynomial
potential. Although this looks like a self-adjoint problem, we will see
that in this formulation the counterterms may in fact become complex.  

\section{The renormalization problem for the pion-less theory}
\label{sec:coor-mom}

The renormalization of the pion-less theory, i.e., a set of pure
contact interactions, has been treated with great detail in the
literature~\cite{Kaplan:1996xu,Cohen:1996my,
Phillips:1996ae,Beane:1997pk,
Phillips:1997xu,Phillips:1998uy,vanKolck:1998bw,Gegelia:1998iu,Ghosh:1998jm,
Kaplan:1999qa, Yang:2004ss,Harada:2005tw,Harada:2006cw} although without much
consideration on how this problem might be embedded into the wider and
certainly more realistic situation where the finite range and short
distance singular chiral NN potentials are present. In fact, much of
the understanding of non-perturbative renormalization within the
modern NN context has been tailored after those and further studies
based on the non-singular OPE singlet $^1S_0$
potential~\cite{Kaplan:1996nv,Kaplan:1999qa,
Gegelia:2001ev,Nieves:2003uu} plus the standard perturbative
experience. In previous~\cite{PavonValderrama:2004nb,
PavonValderrama:2005gu,PavonValderrama:2004td,Valderrama:2005wv,PavonValderrama:2005uj}
and in the present work, we pursue exactly the opposite goal: we will
only consider renormalization procedures which can directly be
implemented in the {\it presence} of long distance potentials since,
after all, contact NN interactions are always assumed to approximate
truly finite range interactions in the long wavelength limit. Thus, it
is useful to review here those developments with an eye on the new
ingredients which appear in the non-perturbative renormalization of
singular pion exchange potentials as analyzed in later sections. In
addition, the deduced running of the counterterms in the contact
theory in the infrared domain serves as a useful starting point when
the long distance pion exchange potential is switched on. Finally, we
will also discuss the size of finite cutoff corrections to the
renormalized result depending both on the particular regularization as
well as the corresponding representation of the short distance
physics.

\subsection{Momentum space} 
\label{sec:mom-pion-less}

Although the previously described momentum space framework has
extensively been used in the past to describe successfully the
data~\cite{Epelbaum:1998ka,Epelbaum:1999dj,Entem:2001cg,
Entem:2002sf,Epelbaum:2003gr,Epelbaum:2003xx,Entem:2003cs,Entem:2003ft,Epelbaum:2004fk}
with a {\it finite cutoff} $\Lambda $ it is worth emphasizing some
puzzling features regarding the off-shell ambiguities of the short
distance potential when finite range corrections, encoded in the
$C_2$, $C_4$ etc. counterterms, are included.

In momentum space, the pion-less theory corresponds to taking $V_L
(k',k) =0$. In such a case the Lippmann-Schwinger equation reduces to
a simple algebraic equation~\cite{Phillips:1997xu,Yang:2004ss}. At
very small values of the cutoff $\Lambda <m_\pi/2 $, the long range
part of the potential may be neglected since they scale with powers of
momentum and a simple contact theory of the form of
Eq.~(\ref{eq:Vs(k,k')}) may be used. For instance, when
$V_S(k',k)=C_0(\Lambda)$, the Lippmann-Schwinger equation (LSE) 
may be directly solved. Using the
basic integral~\footnote{The
result for a different momentum cutoff scheme such as $V (k',k) \to
g(k',\Lambda) V(k,k') g (k, \Lambda )$ corresponds to making the
replacement $\int_0^\Lambda dq \to \int dq g(q,\Lambda)^2 $. In dimensional regularization (minimal subtraction scheme) 
the integral is just the unitarity piece, $ J_0= - i \frac{\pi p}2  $.}
\begin{eqnarray}
J_0=\int_0^\Lambda \frac{d q \, q^2}{p^2-q^2+ i 0^+}= -\Lambda - i \frac{\pi p}2 
+\frac{p}{2}\log \frac{\Lambda+p}{\Lambda-p} \, , 
\end{eqnarray} 
for a sharp momentum cutoff $\Lambda$ and  $0 \le p \le
\Lambda$, one obtains for the phase shifts
\begin{eqnarray}
p \cot \delta (p) &=& -\frac2{M \pi C_0} - \frac{2\Lambda}{\pi} + 
\frac{p}{\pi}\log \frac{\Lambda+p}{\Lambda-p} \, . 
\end{eqnarray} 
At zero energy, $T(0,0) = 2 \alpha_0 / M \pi $ and, thus, the running of $C_0$
is given by 
\begin{eqnarray}
 M \Lambda  C_0 ( \Lambda ) = -\frac{\alpha_0}{\alpha_0 -
 \frac{\pi}{2 \Lambda}} \,\,  .
\label{eq:c0-short}
\end{eqnarray}
 In this case, the phase shift is given by 
\begin{eqnarray}
p \cot \delta (p) &=& - \frac1\alpha_0 +  \frac{p}{\pi}\log
\frac{\Lambda+p}{\Lambda-p} \nonumber \\ 
&=& - \frac1\alpha_0 + {\cal O} 
(\frac1{\Lambda}) \, ,  
\label{eq:pcotdmom0}
\end{eqnarray} 
which corresponds to an ERE with $r_0=v_2= \dots =0$ in the limit
$\Lambda \to \infty$. Note that finite cutoff corrections scale as
$1/\Lambda$. This indicates a relatively slow convergence towards the
infinite cutoff limit and hence that finite cutoff effects are
quantitatively important and might even become a parameter of the
theory. Actually, one might determine $\Lambda$ by fixing the
effective range from the first of Eq.~(\ref{eq:pcotdmom0}), $r_0
(\Lambda)= 4 /(\pi \Lambda) = 2.77\: {\rm fm} $, yielding the accurate
numerical value $\Lambda= 90.7\: {\rm MeV}$. In this case, this particular
three-momentum regularization method becomes itself a model, since we
have no control on the remainder. In any case, it is straightforward
to check that for any finite cutoff there is no off-shellness: $
T(k',k) = T(p,p)$.

The running given by Eq.~(\ref{eq:c0-short}) must be used for {\it
any} cutoff $\Lambda$ if we want to renormalize in the end. However,
thinking of the more general case where finite range corrections are
relevant such a running is only reliable for very small cutoffs
$\Lambda \ll \pi / 2 \alpha_0 $. If we consider also a $C_2(\Lambda)$
coefficient in the potential, the corresponding LSE can be solved with
the ansatz
\begin{eqnarray}
T(k',k) = T_0 (p) + T_2 (p) (k^2+ {k'}^2) + T_4 (p) k^2 {k'}^2 \, , 
\end{eqnarray} 
which yields a set of three linear equations for $T_0(p)$, $T_2(p)$
and $T_4 (p)$. After some algebraic manipulation, the final result for
the phase shift can then be written in the form
\begin{eqnarray}
p \cot \delta (p) &=& \frac{10(C_2 M \Lambda^3 +3)^2 / (M \pi)}{9 ( C_2^2 M \Lambda^5
- 5 C_0)-15 C_2  (C_2 M \Lambda^3 +6)p^2 } \nonumber \\ 
&-& \frac{2\Lambda}{\pi} +
\frac{p}{\pi}\log \frac{\Lambda+p}{\Lambda-p} \, . 
\end{eqnarray} 
Matching at low energies to the ERE, Eq.~(\ref{eq:ERE}), we get the
running of $C_0$ and $C_2$
\begin{eqnarray} 
-\frac1{\alpha_0} &=& \frac{10(C_2 M \Lambda^3 -3)^2}{9 M \pi (
-C_2^2 M \Lambda^5 + 5 C_0)} - \frac{2\Lambda}{\pi} \nonumber \\ \frac12
r_0 &=& \frac{50 C_2 \left( 3 + C_2 M \Lambda^3 \right)^2 \left( 6 + C_2 M 
\Lambda^3\right)}{27 \pi \left( -5 C_0 + C_2^2 \Lambda^5 M \right)^2  }  + \frac{2 }{\pi \Lambda} \, . 
\nonumber \\
\end{eqnarray} 
The first equation allows to eliminate uniquely $C_0$ in favour of
$\alpha_0$ and $C_2$, but as we see there are two branches for the
solutions. However, we choose the branch for which $C_2$ decouples in
the infrared domain, i.e.  fulfills $C_2 \to 0 $ for $\Lambda \to
0$. In fact, at small cutoffs, one gets for this branch
\begin{eqnarray} 
M C_0 (\Lambda) \Lambda &=&  \frac{2 \alpha_0 \Lambda}{\pi} + 
\left( \frac{2 \alpha_0 \Lambda}{\pi} \right)^2 + \frac23  
\left( \frac{2 \alpha_0 \Lambda}{\pi} \right)^3 +  \dots  
\nonumber \\ 
M C_2 (\Lambda) \Lambda^3 &=& -\frac12 \left( \frac{2 \alpha_0 \Lambda}{\pi} \right)^2 +  \dots  \nonumber \\ 
\label{eq:C0-C2}
\end{eqnarray} 
The factor $2/3$ appearing in the small cutoff expansion for
$C_0$ differs already from the coefficient in the case
$C_2=0$. Eliminating $C_0$ and $C_2$ in favour of $\alpha_0$ and
$r_0$, the phase shift becomes
\begin{eqnarray}
p \cot \delta (p) &=& - \frac{2 \Lambda}{\pi \alpha_0} \frac{(\pi - 2
\Lambda \alpha_0)^2}{ 2 \Lambda (\pi - 2 \Lambda \alpha_0) + \alpha_0
p^2 (r_0 \pi \Lambda -4)} \nonumber \\ &-& \frac{2\Lambda}{\pi} +
\frac{p}{\pi}\log \frac{\Lambda+p}{\Lambda-p} \nonumber \\ &=& -
\frac1\alpha_0 + \frac12 r_0 p^2 + {\cal O} (\frac1{\Lambda}) \, . 
\label{eq:ps-Lambda}
\end{eqnarray} 
Note that the finite cutoff corrections are, after fixing $r_0$,
again ${\cal O} ( \Lambda^{-1}) $. So, fixing more low energy
constants in the contact theory does not necessarily imply a stronger
short distance insensitivity, as one might have naturally
expected~\footnote{We have in mind dispersion relations where any
subtraction at zero energy and derivatives thereof of the dispersive
part improve the high energy behaviour and become more insensitive in
the ultraviolet. As we see this is not the case in the contact
pion-less theory.}.  In other words, the inclusion of a higher
dimensional operator such as $C_2$ does not improve the ultraviolet
limit, at least in the polynomial representation given by
Eq.~(\ref{eq:Vs(k,k')}). In Sec.~\ref{sec:N3LO} we will show, however,
that with just one counterterm $C_0$ the inclusion of pion exchange
long distance contributions generates a much faster convergence
towards the renormalized limit as anticipated in
Refs.~\cite{PavonValderrama:2005gu,Valderrama:2005wv,PavonValderrama:2005uj}
(see also Ref.~\cite{Valderrama:2007nu} for a quantitative estimate).
In Sec.~\ref{sec:weinberg} we will also show that when a $C_2 $
counterterm is added this scaling behaviour is not only broken but
also the phase shift fails to converge in the limit $\Lambda \to
\infty$.

Thus, we see that one can establish a one-to-one mapping between the
counterterms $C_0$, $C_2$ and the threshold parameters $\alpha_0$ and
$r_0$. Nevertheless, this is done at the expense of operator mixing,
i.e., both $C_0$ and $C_2$ are intertwined to determine both the
scattering length and the effective range. In other words the cutoff
dependence of $C_0 $ is different depending on the presence of
$C_2$. As we have seen this is not a problem since for small cutoffs
we expect the running of $C_0$ to be fully independent of $C_2$ and
hence on $r_0$. However, unlike the one counterterm case, where
$C_2=0$, the solutions of Eq.~(\ref{eq:C0-C2}) may become complex when
\begin{eqnarray}
\alpha_0^2 r_0 \pi \Lambda^3 - 16 \alpha_0^2 \Lambda^2 + 12 \alpha_0
\pi \Lambda -3 \pi^2 \le 0 \, . 
\end{eqnarray} 
For the physical $^1S_0$ threshold parameters this happens already for
$\Lambda > \Lambda_c = 382 {\rm MeV}$ (the other two roots are
complex). Above this critical value the potential violates
self-adjointness. For $r_0 \to 0$ one has $\Lambda_c \to 16/(\pi r_0)
\to \infty $. Thus, the cutoff can only be fully removed with a
self-adjoint short distance potential if $r_0=0$. This is consistent
with the violation of the Wigner causality condition reported in
\cite{Phillips:1996ae,Beane:1997pk, Phillips:1997xu,Phillips:1998uy}.
Note that the violation of self-adjointness is very peculiar since
once $C_0$ and $C_2$ have been eliminated the phase-shift
(\ref{eq:ps-Lambda}) remains real~\footnote{Nonetheless, off-shell
unitarity deduced from sandwiching the relation $ T-T^\dagger = - 2\pi
i T^\dagger \delta (E-H_0) T$ between off-shell momentum states, is
violated, since the Schwartz's reflection principle fails $\left[T (E+
{\rm i} 0^+) \right]^\dagger \neq T (E-{\rm i} 0^+) $. This would also
have far reaching consequences for the three body problem, since three
body unitarity rests on two-body off-shell unitarity and
self-adjointness of three body forces.}.

One feature in the theory with two counterterms $C_0$ and
$C_2$ is that the off-shell $T-$matrix becomes on-shell only in the
infinite cutoff limit,
\begin{eqnarray}
T(k',k) = T(p,p) + {\cal O} \left( \Lambda^{-1} \right) \, .  
\end{eqnarray} 
This is unlike the theory with one counterterm $C_0$ where there is no
off-shellness at any cutoff. Thus, finite cutoff effects are also a
measure of the off-shellness in this particular problem. This will
have important consequences in Sec.~\ref{sec:weinberg} when attempting
to extend the theory with two counterterms in the presence of the
long distance pion exchange potentials since the off-shellness of the
short distance contribution of the potential becomes an issue in the
limit $\Lambda \to \infty$.

The situation changes qualitatively when the fourth order corrections
depending on two counterterms $C_4 $ and $C_4'$ are
considered. Obviously, we cannot fix both $C_4 $ and $C_4'$ {\it
simultaneously} by fixing the slope parameter $v_2$ of the effective
range expansion, Eq.~(\ref{eq:ERE}). Clearly, one expects some
parameter redundancy between $C_4 $ and $C_4'$ or else an
inconsistency would arise since a sixth order parameter in the
effective range expansion $v_3$ should be fixed. The situation worsens
if higher orders in the momentum expansion are considered due to a
rapid proliferation of counterterms while there is only one more
threshold parameter for each additional order in the expansion. This
required parameter redundancy is actually a necessary condition for
consistency which is manifestly fulfilled within dimensional
regularization but not in the three-momentum cutoff
method~\footnote{This operator redundancy has also been discussed on a
Lagrangean level~\cite{Beane:2000fi} based on equations of motion and
in the absence of long distance interactions ( see
also~\cite{Barford:2002je}).}. Moreover, it was realized some time
ago~\cite{Phillips:1997xu,Beane:1997pk} that the finite cutoff
regularization and dimensional regularizarion in the minimal
subtraction scheme yielded {\it different} renormalized amplitudes for
a {\it truncated} potential. This non-uniqueness in the result due to
a different regularization happens when a non-vanishing $C_2$
counterterm is considered. In any case, the dimensional regularization
scheme has never been extended to include the long range part of the
TPE potential which usually appear in the present NN context. Thus,
for the momentum space cutoffs which have been implemented in
practice the short distance representation is somewhat inconsistent at
least for a finite value of the cutoff $\Lambda$.

Alternatively, one may choose an energy dependent representation of
the short distance physics as 
\begin{eqnarray}
V_S = C_0 + 2 p^2 C_2 + p^4 ( 2C_4 + C_4') + \dots 
\label{eq:Vs(p)}
\end{eqnarray}  
In this case the correspondence between counterterms and threshold
parameters $\alpha_0$,$r_0$, $v_2$, etc. is exactly one-to-one, and
the parameter redundancy is manifest, since the on shell $T$-matrix
depends only on the on-shell potential. Actually, under dimensional
regularization the representations of the potential
Eq.~(\ref{eq:Vs(k,k')}) and Eq.~(\ref{eq:Vs(p)}) yield the same
scattering amplitude.  Although this on-shell equivalence is certainly
desirable it is also unnatural, if the long distance potential is
energy independent. We will nevertheless analyze such a situation in
the next subsection in coordinate space.

The previous discussion highlights the kind of undesirable but
inherent off-shell ambiguities which arise when finite range
corrections are included in the short distance
potential~\footnote{This fact becomes more puzzling if the potential $
V= C_2 ( k^2 + {k'}^2 - 2 p^2) $ is considered. It vanishes on the mass
shell $k=k'=p$ but nonetheless generates non trivial on shell
scattering for the three-momentum cutoff.}. In our view these are
unphysical ambiguities which have nothing to do with the unambiguous
off-shell dependence of the long distance potential. Of course, one
way to get rid of the ambiguities is to take the limit $\Lambda \to
\infty$ which corresponds to the case where a truly zero range theory
is approached.  However, even for a finite cutoff there is a case
where one is free from the ambiguities, namely when the short distance
potential is {\it both} energy and momentum independent for s-wave
scattering
\begin{eqnarray}
V_S (k',k) = C_0 (\Lambda)  \, .
\end{eqnarray} 
The key point is that we allow only this counterterm to be cutoff
dependent and real, as required by self-adjointness.  Of course, the
discussion above for the contact theory suggests the benefits of using
just one $C_0$ counterterm but does not exactly provide a proof that
one {\it must} take further counterterms such as $C_2$ to zero.  The
extension of this analysis to the case of singular chiral potentials
in Sec.~\ref{sec:weinberg} will yield the definite conclusion that
renormalizability is indeed equivalent to take $C_2=0$.

\subsection{Coordinate space} 
\label{sec:coor-pion-less}

The previous renormalization scheme is the momentum space version
corresponding to the coordinate space renormalization adopted in a
previous work by two of us (MPV and
ERA)~\cite{PavonValderrama:2005gu,Valderrama:2005wv,PavonValderrama:2005uj}.
Actually, in the pure contact theory, we can relate the renormalization
constant with the momentum space wave function explicitly. At large
values of the short distance cutoff $r_c$, the zero energy wave
function reads,
\begin{eqnarray}
u_0 (r_c) = 1 - \frac{r_c}{\alpha_0} \, .  
\end{eqnarray} 
Thus, the following relation holds  
\begin{eqnarray} 
\frac{\alpha_0}{\alpha_0 -r_c} = 1 - r_c \frac{u_0'(r_c)}{u_0 (r_c)} \, .  
\end{eqnarray} 
Comparing with Eq.~(\ref{eq:c0-short}), we get 
\begin{eqnarray} 
M \Lambda  C_0 ( \Lambda ) = r_c \frac{u_0'(r_c)}{u_0 (r_c)} \Big|_{r_c= \pi / 2 \Lambda} - 1 \, , 
\label{eq:C0-equiv}
\end{eqnarray} 
where the momentum cutoff $\Lambda$ and the short distance cutoff
$r_c$ are related by the equation  
\begin{eqnarray} 
\Lambda r_c = \frac{\pi}{2}
\label{eq:Lambda-rc} 
\end{eqnarray} 
which is nothing but an uncertainty principle relation between
cutoffs~\footnote{This relation will be shown to hold also in the
presence of a local potential, see Appendix~\ref{sec:nyquist}.}.  Note
that for the standard regular solution $u_0 (r) \sim r $ one has a
vanishing counterterm $ C_0 =0 $. In contrast, $ C_0 \neq 0 $ for the
irregular solution. In the case of the singular attractive potentials
the solution is regular but highly oscillatory and the $C_0 $ takes
all possible values for $r_c \to 0$.  A more detailed discussion on
these issues can be seen in
Refs.~\cite{PavonValderrama:2003np,PavonValderrama:2004nb,PavonValderrama:2004td}. Of
course, strictly speaking both Eq.~(\ref{eq:C0-equiv}) and
Eq.~(\ref{eq:Lambda-rc}) are based on a zero energy state, and in the
finite energy case we will assume these relations having the limit
$r_c \to 0$ or $\Lambda \to \infty $ in mind.

Let us now deal with finite energy scattering states. Since there is
no potential, $U_L(r)=0$, for $r > r_c$ we have the free wave solution
\begin{eqnarray} 
u_p (r) = \frac{\sin (pr + \delta (p)) }{\sin \delta(p)} \, .  
\end{eqnarray} 
In the theory with one counterterm, we fix the scattering length
$\alpha_0$ by using the zero energy wave function and matching at
$r=r_c$ so we get 
\begin{eqnarray} 
\frac{u_0' (r_c)}{u_0 (r_c)} = \frac{u_p' (r_c)}{u_p (r_c)} = p \cot
(p r_c + \delta (p)) \, ,
\end{eqnarray} 
yielding 
\begin{eqnarray} 
\frac1{r_c- \alpha_0 }= p \cot (p r_c + \delta (p)) \, , 
\end{eqnarray} 
and thus 
\begin{eqnarray} 
p \cot \delta (p) &=& -p \,  \frac{1- p ( \alpha_0 -r_c ) \tan (p r_c)}{p (
 \alpha_0 -r_c) + \tan (p r_c) }  \, \nonumber \\
&=& - \frac1{\alpha_0} + {\cal O} (r_c) \, .   
\end{eqnarray} 
This is in qualitative agreement with the momentum space result when
the cutoff is being removed, Eq.~(\ref{eq:pcotdmom0}) and, as we can
see, the approach to the renormalized value is similar if the
identification $r_c = \pi / (2 \Lambda )$ is made. Note further that
since the boundary condition is energy independent the problem is
self-adjoint and hence orthogonality between different energy states
is guaranteed.

The theory with two counterterms where both $\alpha_0$ and $r_0$ are
fixed to their experimental values opens up a new possibility, already
envisaged in Ref.~\cite{Valderrama:2007nu}, related to the
non-uniqueness of the result both for a finite cutoff as well as for
the renormalized phase-shift. As pointed out above, this
non-uniqueness was noted first in momentum space
Ref.~\cite{Phillips:1997xu,Beane:1997pk} when using a finite three
dimensional cutoff or dimensional regularizarion (minimal
subtraction). Remarkably, within the boundary condition regularization
we will be able to identify both cases as different short distance
representations. 

Actually, when fixing $\alpha_0$ and $r_0$ we are led to
\begin{eqnarray}
\frac{u'_p (r_c)}{u_p(r_c)} = d_p (r_c) = \frac{u'_0 (r_c) + p^2 u'_2 (r_c)}{ u_0(r_c)+ p^2 u_2(r_c)} + {\cal O} (p^4)  \, . 
\end{eqnarray} 
where $u_0$ and $u_2$ are defined in Sec.~\ref{sec:coor-space}. Note
that now self-adjointness is violated from the beginning due to the
energy dependence of the boudary condition.  Within the second order
approximation in the energy the neglected terms are ${\cal O} (p^4)$,
so {\it any} representation compatible to this order might in
principle be considered as equally suitable. The close similarity to a
Pad\'e approximant suggests to compare the following three
possibilities for illustration purposes
\begin{eqnarray}
d_p^{\rm I} &=& \frac{u'_0 (r_c) + p^2 u'_2 (r_c)}{ u_0(r_c)+ p^2
u_2(r_c)} \, , \nonumber \\ d_p^{\rm II} &=& \frac{u_0'(r_c)}{u_0(r_c)} +
p^2 \left[ \frac{u_2'(r_c)}{u_0(r_c)} - \frac{u_0'(r_c)
u_2(r_c)}{u_0(r_c)^2} \right] \, , \nonumber \\ d_p^{\rm III} &=&
\frac{u_0' (r_c)^2}{u_0 (r_c) u_0'(r_c) + p^2 \left[ u_2
(r_c)u_0'(r_c) - u_0 (r_c) u_2' (r_c) \right] } \, , \nonumber \\ 
\label{eq:d_p's}
\end{eqnarray} 
and study what happens as the cutoff is removed, $r_c \to 0$. Note
that all three cases possess {\it by construction} the same scattering
length $\alpha_0$ and effective range $r_0$ and no potential for $r>r_c$. 
Straightforward calculation yields 
\begin{eqnarray} 
p \cot \delta (p) &=& -\frac1{\alpha_0} + \frac12 r_0 p^2 + {\cal O} (r_c^2)
\qquad {\rm (I)} \nonumber \\ p \cot \delta (p) &=& -\frac1{\alpha_0} + \frac12
r_0 p^2 + {\cal O} (r_c) \qquad {\rm (II)} \nonumber \\ p \cot
\delta (p) &=& -\frac1{\alpha_0} \frac1{ 1 - \frac12 \alpha_0 r_0 p^2}
+ {\cal O} (r_c) \qquad {\rm (III)} \nonumber \\
\label{eq:coor-pion-renor}
\end{eqnarray}
As we see, all three representations provide the {\it same} threshold
parameters, but {\it do not} yield identical renormalized amplitude
for finite energy. Actually, cases I and II coincide with the
three-dimensional cutoff regularization method (see
Sec.~\ref{sec:mom-pion-less}), whereas case III corresponds to
dimensional regularization (MS).  Moreover, the finite cutoff
corrections to the renormalized result are, generally, ${\cal O} (r_c)
$ while the rational representation yields corrections ${\cal O}
(r_c^2) $. These observations survive at higher orders when $v_2$,
$v_3$, etc.\ threshold parameters are further taken into account. This
indicates that not all short distance representations are equally
``soft'' in the UV-cutoff. The generalization of these results to the
case of singular TPE chiral potentials was studied in
Ref.~\cite{Valderrama:2007nu} and will be also re-analyzed in
Sec.~\ref{sec:weinberg} while discussing the consistency of the
standard Weinberg's power counting.

In any case, when finite range corrections are considered within the
boundary condition regularization, there are two possible renormalized
solutions depending on the particular parameterization of short distance
physics.  A nice feature of this regularization is that they can be
identified with similar results found already in the momentum space
analysis of Ref.~\cite{Phillips:1997xu,Beane:1997pk} when confronting
three-momentum cut-off and dimensional regularization. We note also
here that no ambiguity arises when the boundary condition is assumed
to be energy independent, in which case self-adjointness is
guaranteed.

\begin{figure}[ttt]
\begin{center}
\epsfig{figure=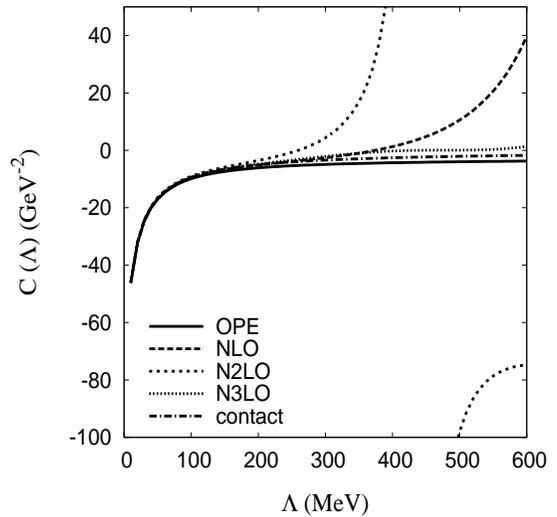,height=7.5cm,width=10cm}
\end{center}
\caption{LO, NLO, N2LO and N3LO running of the counterterm (in ${\rm
GeV}^{-2}$ as a function of the cutoff $\Lambda$ in the $^1S_0$ channel
for small cutoffs $\Lambda \le 600 {\rm MeV}$. The renormalization
condition is determined by fixing the scattering length to its
experimental value $\alpha_0 = -23.74 {\rm fm}$. We use the parameters
of Ref.~\cite{Entem:2003ft} for the pion exchange potential $V_L$.}
\label{fig:c0-1S0.eps}
\end{figure}

\begin{figure*}[ttt]
\begin{center}
\epsfig{figure=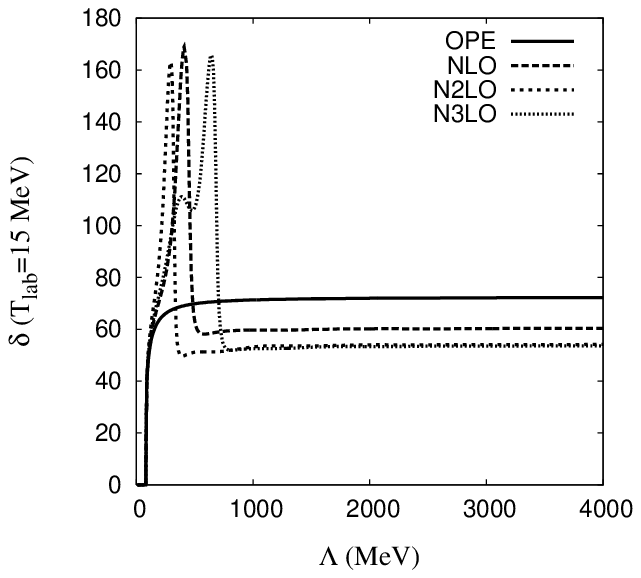,height=4.5cm,width=8.5cm}
\epsfig{figure=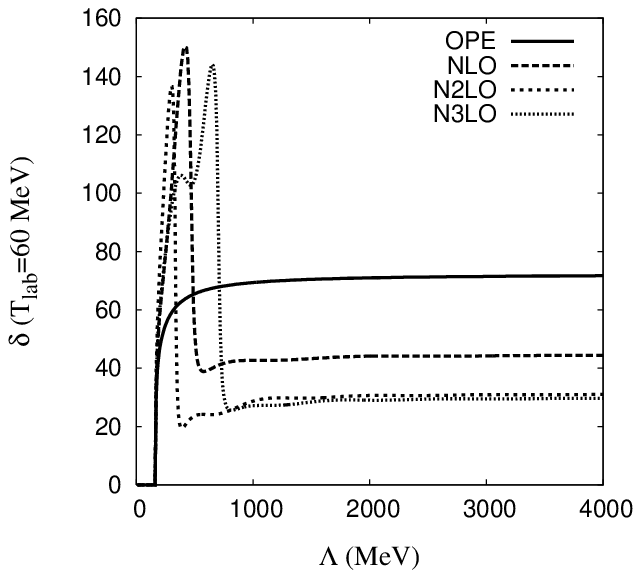,height=4.5cm,width=8.5cm}\\ 
\epsfig{figure=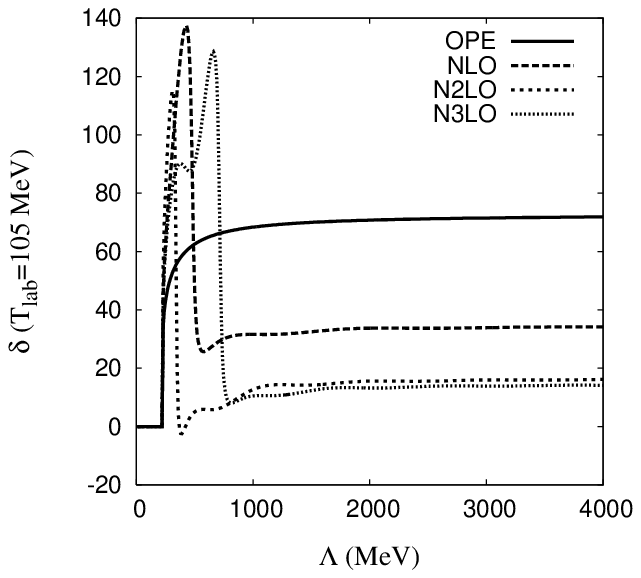,height=4.5cm,width=8.5cm} 
\epsfig{figure=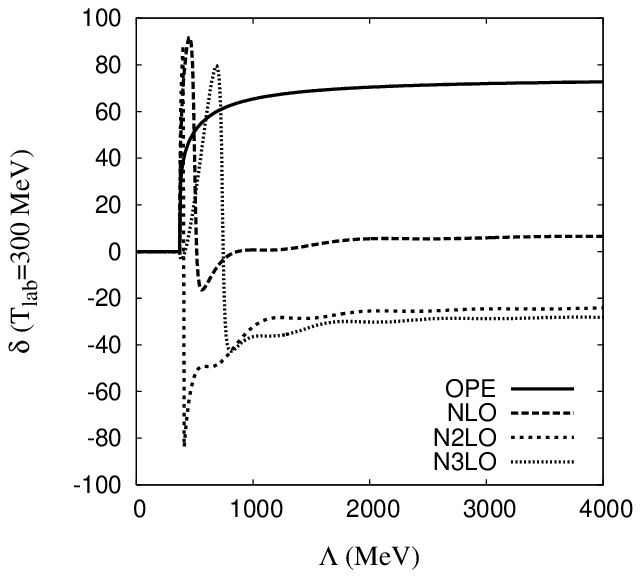,height=4.5cm,width=8.5cm}
\end{center}
\caption{LO, NLO, N2LO, and N3LO convergence of the phase shifts as a
function of the momentum cutoff $\Lambda $ for fixed LAB energies,
$T_{\rm LAB}=15,60,105,300\: {\rm MeV}$.  The renormalization
counterterm $C_0 (\Lambda) $ is always determined by fixing the scattering
length to its experimental value $\alpha_0 = -23.74 {\rm fm}$.}
\label{fig:1S0-cutoff.eps}
\end{figure*}

\section{Renormalization of Pion Exchanges with One Counterterm} 
\label{sec:N3LO} 

The study of the contact theory in Sec.~\ref{sec:coor-mom} provides
suggestive arguments why it is highly desirable to carry out a
regularization with a single counterterm in the $^1S_0$ channel by
adjusting it to the physical scattering length for any cutoff
value. In this section we want to extend that study when the long
distance chiral potential organized according to Weinberg power
counting enters the game and the cutoff is removed. By taking the
cutoff to infinity, we are actually
assuming that all degrees of freedom not included in the
present calculation become infinitely heavy. This way we expect to
learn about missing physics in a {\it model} and {\it
regularization} independent fashion. The traditional strategy of
adjusting an increasing number of counterterms 
may obscure the analysis.
In other words, by using this minimal number of counterterms,
we try not to mock up what might be still missing in
the long range description.

In this context there is of course the question of convergence or
cutoff insensitivity of the phase shift, when $\Lambda \to \infty $,
provided we keep at any rate the scattering length $\alpha_0$ to its
physical value by suitably adjusting the unique counterterm $C_0
(\Lambda) $.  In coordinate space this is a rather trivial matter if
the long distance potential is
local~\cite{PavonValderrama:2005gu,Valderrama:2005wv,PavonValderrama:2005uj},
as it happens in the LO, NLO, and N2LO Weinberg counting. The analysis
in momentum space involves detailed large momenta behaviour of the
Lippmann-Schwinger equation and one must resort to trial and
error. Indeed, the N3LO case analyzed below includes nonlocalities and
it {\it turns out} to provide convergent results.

For numerical calculations, we take the values for the $c_i$ and $d_i$
parameters appearing in the pion exchange potential $V_\pi$ used
in Ref.~\cite{Entem:2003ft}, which do a good job for peripheral
waves, where re-scattering effects are suppressed and where one is,
thus, rather insensitive to cutoff effects. We will only consider TPE
contributions to the N3LO potential.

\subsection{Renormalized N3LO-TPE}

To determine the running of the counterterm we start from low cutoffs
$\Lambda \ll m_\pi $ since the long range part of the potential is
suppressed and Eq.~(\ref{eq:c0-short}) may be used.  Actually, the
analytical result is well reproduced by the numerical method used to
solve the Lippmann-Schwinger equation in the pure contact theory with
$C_0$. Once this identification has been done, the value of the cutoff
is increased steadily so that the scattering length is always fixed to
the experimental value, $\alpha_0 = -23.74 {\rm fm}$. This adiabatic
switching on of the long range physics guarantees that we are always
sitting on the correct branch which smoothly goes into the contact
theory at low cutoffs~\footnote{In general, there may appear many
solutions for $C_0 (\Lambda)$ fitting $\alpha_0$. They are physically
unacceptable unless they behave as $ M C_0 (\Lambda) \to 2 \alpha_0
/\pi $ for $\Lambda \to 0$ since they do not evolve into the theory
where the long range components are decoupled.}.

The running of the counterterm $C_0 (\Lambda) $ is depicted for the
LO, NLO, N2LO, and N3LO potentials in Fig.~\ref{fig:c0-1S0.eps} in the
low cutoff region. As expected, the deviations from the simple result
of the pure contact theory, Eq.~(\ref{eq:c0-short}), start at $\Lambda
\sim m_\pi$ for LO because of the $1\pi$ exchange potential. For
higher cutoffs the NLO, N2LO, and N3LO counterterm displays a cycle
structure very similar to what has been observed in coordinate
space~\cite{PavonValderrama:2004td}.

The convergence of the $^1S_0 $ phase shift for fixed values of the Lab
energy, $T_{\rm LAB}=15,60,105,300\: {\rm MeV}$ is displayed in
Fig.~\ref{fig:1S0-cutoff.eps}. Of course, one observes a faster
convergence for small energies. For the maximal value of $T_{\rm
LAB}=300\: {\rm MeV}$, cutoff values $\Lambda \sim 2\: {\rm
GeV}$ are needed to change the phase shift by less than $1^o$.

In agreement with the analytical estimates of
Ref.~\cite{Valderrama:2007nu}, the convergence of the regulated
phase-shifts towards their renormalized values follows a computable
power like pattern, $\delta (k) - \delta (k, \Lambda) = {\cal O} (
\Lambda^{-n/2-1}) $.  The more singular the potential at large momenta
the faster the convergence. Thus, the expected increased insensitivity
at short distances is indeed confirmed.

\begin{figure}[ttt]
\begin{center}
\epsfig{figure=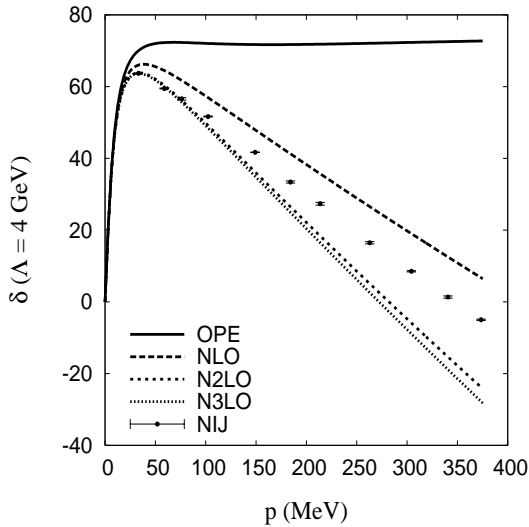,height=7.5cm,width=10cm}
\end{center}
\caption{LO, NLO, N2LO, and N3LO renormalized phase shifts in the
$^1S_0$ channel as a function of the CM momentum compared to the
Nijmegen Partial Wave Analysis~\cite{Stoks:1993tb}. Only one
counterterm is used and fixed by the physical value of the scattering
length.}
\label{fig:ps1S0.eps}
\end{figure}

Finally, the renormalized $^1S_0$ phase shift is presented in
Fig.~\ref{fig:ps1S0.eps} for LO, NLO, N2LO, and N3LO. As a check of the
present calculation in momentum space, let us mention that we reproduce
the coordinate space renormalized phase shifts of
\cite{Valderrama:2005wv} at LO, NLO, and N2LO. For instance, at the
maximal CM momentum of $p=400\: {\rm MeV}$, the maximal discrepancy
between the coordinate space and momentum space phase shifts is less
than half a degree when $r_c = 0.1 {\rm fm}$ and $\Lambda=4\: {\rm GeV}$,
respectively.
  
The clear converging pattern can be observed all over the elastic
scattering region, actually at $T_{\rm LAB} = 300\: {\rm MeV}$, one has
$\delta^{\rm LO} = 72.72^o$ , $\delta^{\rm NLO} = 6.44^o $, $\delta^{\rm
N2LO} = -24.20^o $, and $\delta^{\rm N3LO} = -28.20^o $.  However, there is
still a discrepancy with the Nijmegen PWA result which at this energy
is $\delta^{\rm Nijm} = -4.68\pm 0.55^o $ for np scattering. 

To have an idea on the uncertainty of the calculated phase shift, we
vary the scattering length $\alpha_0 = -23.74(2)$, the value of,
$g_{\pi NN}=13.1(1)$ and $g_A=1.26(1)$. Actually the error in the
latter is correlated through the Goldberger-Treiman relation, so we
will quote both as an error in $g_A$ only. In addition, for the chiral
constants we take the central values used in \cite{Entem:2003ft} which
provided a good description for the peripheral waves and for the
uncertainties we assume as an educated guess those from the $\pi N$
study~\cite{Buettiker:1999ap}.  The only exception is $c_4$, for which
the error from $\pi N$~\cite{Buettiker:1999ap} is much smaller than
the systematic discrepancy with the NN
determination~\cite{Entem:2003ft}. So it is more realistic to take the
systematic discrepancy as the error. Thus we take $c_1=-0.81(15)$,
$c_2=2.80(23)$, $c_3=-3.20\pm1.35$, $c_4=4.40\pm 1.0$,
$d_1+d_2=3.06(21)$,$ d_3=-3.27(73)$, $d_{14}-d_{15}=-5.65(41)$. The
results for the particular variations are presented in table~
\ref{tab:table1} and as we clearly see, the uncertainty in $c_3$
dominates the total error. At $T_{\rm LAB} = 300\: {\rm MeV}$, one has
$\delta^{\rm LO} =75.90 \pm 0.2^o$, $\delta^{\rm NLO}=6.5 \pm 0.7^o$,
$\delta^{\rm N2LO} =-24 \pm 6^0$, $\delta^{\rm N3LO} = -28 \pm 9^o
$. As we see the statistical uncertainty stemming from the input
parameters increases with the order. If we take the difference
$\delta^{\rm N3LO}-\delta^{\rm N2LO} \sim 4^o$ as an estimate of the
systematic error, adding in quadrature statistical and systematic
errors we have, $\Delta \delta^{\rm N3LO}_{\rm TOT} \sim 10^o$, still
a smaller quantity than the discrepancy to the Nijmegen phase shift.
This may suggest that {\it after renormalization} there are some
physical effects missing even at the N3LO level beyond pure TPE.

\begin{table*}
\begin{tabular}{|c|c|c|c|c|c|c|c|c|c|c|c|}
\hline 
 & 
$\Delta \alpha_0 $ & 
$ \Delta g_A $ & 
$\Delta c_1 $ & 
$\Delta c_2 $ & 
$\Delta c_3 $ & 
$\Delta c_4 $ & 
$\Delta (d_1+d_2) $ & 
$\Delta d_3 $ & 
$\Delta d_5 $ & 
$\Delta (d_{14}-d_{15}) $ & 
TOTAL \\ 
\hline 
$\Delta \delta^{\rm LO}$ &  0 & 0.2 & -- & -- & -- & -- & -- & -- & -- & -- & 0.2  \\
\hline 
$\Delta \delta^{\rm NLO}$ &  0 & 0.7 & -- & -- & -- & -- & -- & -- & -- & -- & 0.7\\

\hline 
$\Delta \delta^{\rm N2LO}$ &  0 & 0.4 & 0.1 & -- & 6 & 1.3 & -- & -- & -- & -- & 6.1  \\
\hline 
$\Delta \delta^{\rm N3LO}$ &  0 & 0.2&  0.2 & 0.2 & 9 & 1.1 & 0.2 & 0.1 & 0 & 0.2 & 9.1 \\
\hline 
\end{tabular}
\caption{\label{tab:table1}
Induced errors in the $^1S_0$ phase shift
(in degrees) at $T_{\rm LAB}=300$ when the input parameters are
varied. The sign ``--'' means that there is no contribution to the
variation and a zero, ``$0$'', means that the change is $\Delta \delta
< 0.1 $. The total result is obtained by summing the partial
contributions in quadrature, $\Delta \delta_{\rm TOTAL} = \sqrt{\sum_i
(\Delta \delta_i)^2 }$}
\end{table*}

\medskip

\subsection{Inclusion of Delta}

In the previous section we have seen that all TPE effects included to
N3LO display a convergent pattern after renormalization, but there is
still some missing physics. Note that while at LO and NLO
the only parameters are $g_A$, $m_\pi$, $M_N$ and
$f_\pi$, at N2LO and N3LO there appear new low energy constants (the $c_i$
and $d_i$, respectively) which can be related to $\pi N$ scattering and
encode short range physics not considered explicitly. The $\Delta$
resonance is an outstanding feature of $\pi N $ scattering and
explains a great deal of the low energy constants. Thus, it is
interesting to analyze the role of explicit $\Delta$ excitations as
intermediate states in the $NN$ potential. The importance of explicit
$\Delta$ degrees of freedom has been emphasized on power counting
grounds in several previous works with finite cutoffs where the
$N\Delta$ splitting is regarded as a small parameter $\sim
m_\pi$~\cite{Ordonez:1995rz,Epelbaum:1999dj,Pandharipande:2005sx,Nogga:2005hy}. The
crucial role played in the renormalization problem has been stressed
in Ref.~\cite{Valderrama:2005wv}. In this section, we
analyze the NN potential with the NLO terms of \cite{Kaiser:1997mw}
together with the $1\Delta$ and $2 \Delta$ in the box diagrams as
computed in Ref.~\cite{Kaiser:1998wa}.  One advantage of such an
approach is that, as compared to the standard $\Delta$-less theory,
there only appears the $N\Delta$ splitting as a parameter.

\begin{figure}[ttt]
\begin{center}
\epsfig{figure=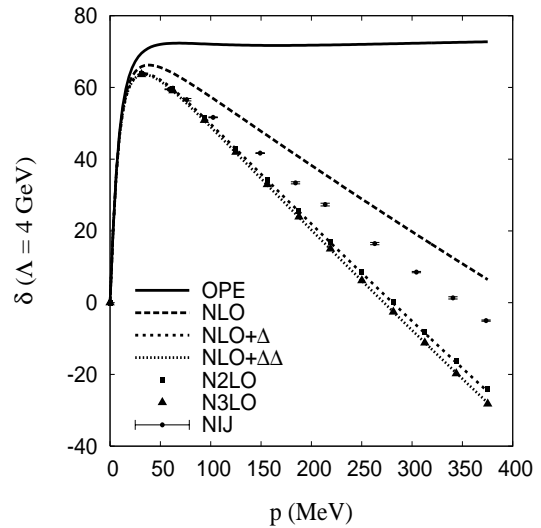,height=7.5cm,width=10cm}
\end{center}
\caption{LO, NLO, NLO+$1\Delta$, and NLO+$2\Delta$ renormalized phase
shifts in the $^1S_0$ channel as a function of the CM momentum
compared to the Nijmegen Partial Wave Analysis~\cite{Stoks:1993tb}.
Only one counterterm is used and fixed by the physical value of the
scattering length.}
\label{fig:ps1S0-del.eps}
\end{figure}

The renormalization of the $^1S_0$ channel proceeds along the lines
discussed in the $\Delta$-less theory. The results for LO, NLO,
NLO+$1\Delta$, and NLO+$2\Delta$ renormalized phase shifts in the
$^1S_0$ channel are plotted and compared to the Nijmegen Partial Wave
Analysis~\cite{Stoks:1993tb} as a function of the CM momentum in
Fig.~\ref{fig:ps1S0-del.eps}. The most striking result is the very
strong resemblance between the N2LO $\Delta$-less vs NLO+$1\Delta$ and
the N3LO $\Delta$-less vs NLO+$2\Delta$. These results sustain the
treatment of the $N \Delta$ splitting as a small parameter $\sim m_\pi
$ corresponding to the $\Delta$-counting NLO$_\Delta$ = NLO +
$1\Delta$+ $2\Delta$.  The fact that all the contributions fall off at
large distances as $\sim e^{- 2 m_\pi r}$ suggests that despite the
ability to mimic higher order corrections in the Weinberg counting for
the long distance potential in the $^1S_0$ channel there is still
missing shorter range physics beyond TPE.

\subsection{Three-pion exchange contributions}

At N3LO, three-pion exchange occurs for the first time.  These
contributions have been calculated by
Kaiser~\cite{Kaiser:1999ff,Kaiser:1999jg} and found to be small, which
is why present N3LO NN potentials omit these contributions when
renormalization is not implemented. However, it should be noted that,
for small distances, the $3\pi$ diagrams are proportional to $r^{-7}$
and, thus, will ultimately dominate at short distances.  A rough
estimate of the results published in
Refs.~\cite{Kaiser:1999ff,Kaiser:1999jg} suggests that the sum of all
$3\pi$ graphs is attractive.  Thus, we infer from there that the
scattering length will still be a free input parameter. On the other
hand the $3 \pi$-exchange contribution falls off as $\sim e^{-3 m_\pi
r}$ at long distances so it becomes active at rather short distances,
and so the effect is expected to be small because of the short
distance suppression (modulo oscillations) of the wave function $u (r)
\sim r^{7/4}$ typical of potentials with a short distance power like
singularity~\cite{PavonValderrama:2005gu,Valderrama:2005wv,PavonValderrama:2005uj}. This
agrees with the rule that the more singular the potential the more
convergent is the calculation, as we have extensively discussed above.
Therefore, in a complete and renormalized N3LO calculation of the
$^1S_0$ phase shifts, we expect the $3\pi$ effects not to be large,
although the predictions may be closer to the empirical values than in
Fig.~\ref{fig:ps1S0.eps}. An accurate investigation of the impact of
$3\pi$ exchange at N3LO on NN phase shifts represents an interesting
and challenging project for the future.

\subsection{Irrelevance of a fixed $C_2$ counterterm}

As we have mentioned, the very definition of the potential is
ambiguous as it requires fixing the polynomial terms in the momentum.
In the calculations presented above, we have taken the renormalization
scheme for the potential where the fixed and cut-off independent
choice $C_2=0$ for the potential is made. We have also analyzed the
situation when a different renormalization scheme is taken, namely a
non-vanishing arbitrary $C_2$ coefficient which does not run with
$\Lambda$. However, we allow $C_0 (\Lambda) $ to run in a way that the
scattering length $\alpha_0 $ is reproduced. This generates a
different renormalization trajectory for $C_0 (\Lambda)$ as compared
to the case $C_2=0$. We find by actual calculations that this fixed
$C_2$ coefficient is irrelevant, i.e. the renormalized phase shift
does not depend on this fixed value in the limit $\Lambda \to
\infty$. Roughly speaking, the reason is that while the polynomial
combination $C_2 q^2 $ is large at large momenta, the pion exchange
part behaves as $q^2\, \log \, (q^2) $ with an additional logarithm
and thus dominates for fixed $C_2$. This irrelevance of $C_2$ was
highlighted in the coordinate space analysis of
Ref.~\cite{Valderrama:2005wv} where the regularization based on a
radial cut-off $r_c$ would provide a compact support not sensing the
details of the distributional contributions in
Eq.~(\ref{eq:pot-coor}), regardless on how many derivatives of delta's
are included. A quite different situation arises when $C_2$ depends on
the cut-off in a way that the effective range is fixed as we discuss
in the next Section~\ref{sec:weinberg}. There, it will be shown that
if $C_2 (\Lambda)$ is relevant then the phase shift is not convergent.

\medskip

\section{Weinberg's counting and renormalization with two counterterms.}
\label{sec:weinberg} 

\subsection{Momentum space}

In previous sections, we have seen that the LO, NLO, N2LO, and N3LO
chiral potentials can be renormalized when one counterterm $C_0
(\Lambda) $ is determined for any value of the cutoff $\Lambda$ by
fixing the scattering length to its experimental value, $\alpha_0 =
-23.74 {\rm fm}$ and the limit $\Lambda \to \infty $ is subsequently
taken.  This agrees with the observation of
Refs.~\cite{PavonValderrama:2005gu,Valderrama:2005wv,PavonValderrama:2005uj}
that attractive singular potentials can be renormalized with a single
counterterm. However, Weinberg counting requires further
counterterms in the short distance potential, see
Eq.~(\ref{eq:Vs(k,k')}).  For instance, both at NLO and N2LO a $C_2 $
counterterm should be included.

In this section, we discuss whether the $^1S_0$ scattering amplitude is
renormalizable, i.e. whether the scattering amplitude has a well
defined limit, when both the $C_0 (\Lambda) $ and $C_2 (\Lambda) $
counterterms are included and fixed by fitting the scattering length
$\alpha_0 = -23.74\: {\rm fm}$ as well as the effective range $r_0= 2.77
\: {\rm fm}$ for any value of the cutoff $\Lambda$ and the limit
$\Lambda \to \infty$ is pursued. Actually, when the NLO (N2LO) long
distance potential, Eq.~(\ref{eq:1pi2pi3pi}) is used, this way of
proceeding corresponds to renormalizing the NLO (N2LO) approximation of
the $^1S_0$ channel in the standard Weinberg counting. The method we
follow in practice is a straightforward extension of the case with
only just one counterterm $C_0$. In the case when $C_2 =0 $, 
we get a finite (renormalized) effective range $r_0^{\rm
NLO}=2.29\: {\rm fm}$, which is close but still differs significantly
from the experimental value. At N2LO with $C_2=0$ one gets $r_0^{\rm
N2LO}=2.86\: {\rm fm}$.  This last value is so close to the experimental
one that one would not expect big changes when a $C_2$ is added to
exactly fit the experimental $r_0$.  Thus, it makes sense to
investigate what would happen if one uses a non-vanishing $C_2$ to
account for the missing, and in principle tiny, contribution to the
effective range. The surprissing result, to be discussed below in
detail, is that trying to fit the discrepancy in the effective range
with a $C_2$ counterterm is incompatible with renormalizability.

\begin{figure}[tbc]
\begin{center}
\epsfig{figure=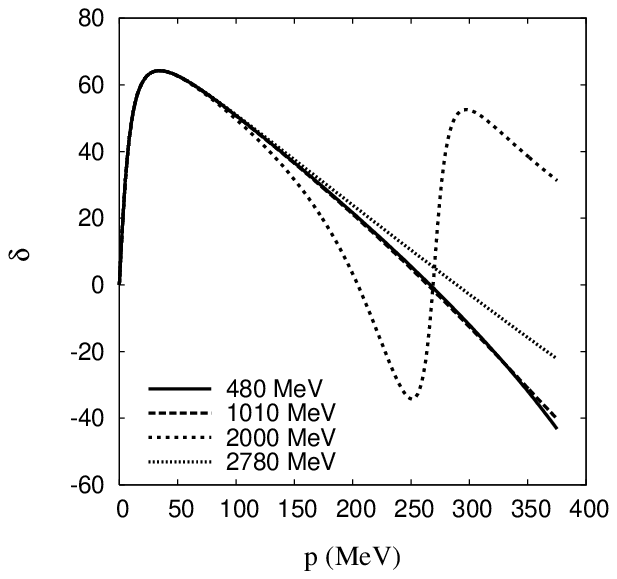,height=6.5cm,width=8.5cm}
\epsfig{figure=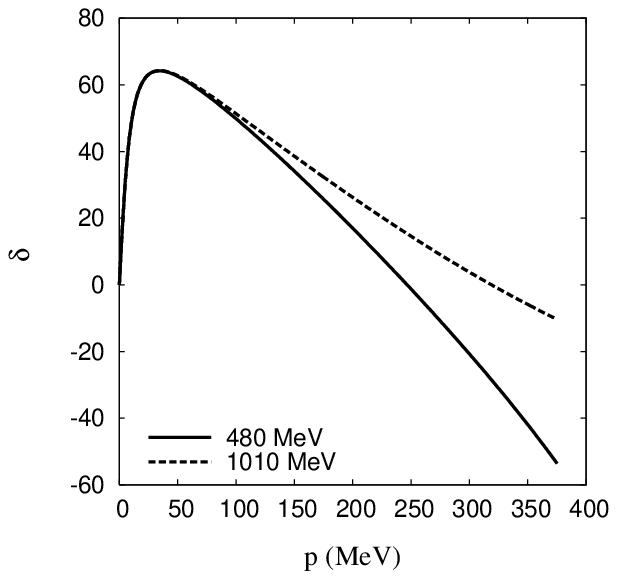,height=6.5cm,width=8.5cm}
\end{center}
\caption{NLO (upper pannel) and N2LO (lower pannel) convergence of
the phase shifts as a function of the CM momentum for some fixed values of
the cutoff $\Lambda $.  The renormalization counterterms $C_0
(\Lambda) $ and $C_2 (\Lambda) $ are always determined by fixing the
scattering length $\alpha_0 $ and the effective range $r_0 $ to their
experimental values $\alpha_0 = -23.74\: {\rm fm}$ and $r_0 = 2.77\: {\rm
fm}$. For $\Lambda > \Lambda_c \sim 500 {\rm MeV}$ both $C_0 (\Lambda)
$ and $C_2 (\Lambda) $ become complex while the phase shifts remain
real.}
\label{fig:1S0-ps-C0-C2-mom.eps}
\end{figure}

\begin{figure}[tbc]
\begin{center}
\epsfig{figure=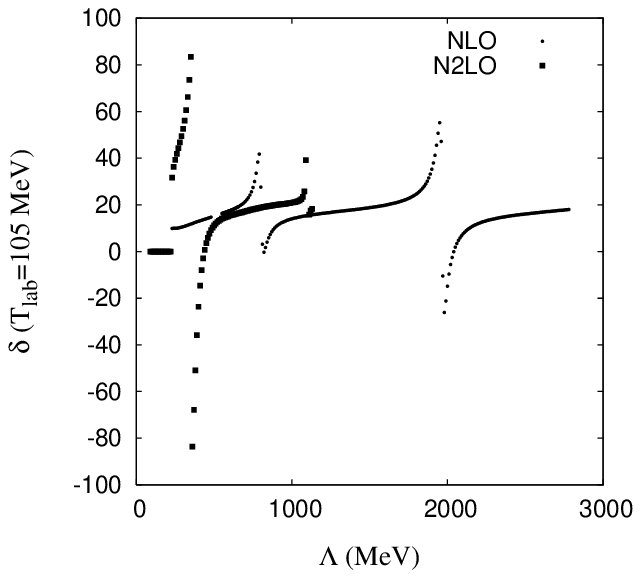,height=6.5cm,width=8.5cm}
\epsfig{figure=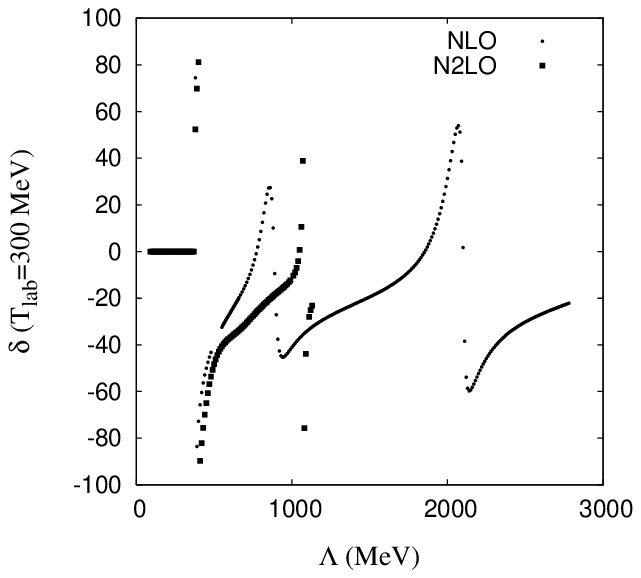,height=6.5cm,width=8.5cm}
\end{center}
\caption{NLO and N2LO convergence of the phase shifts as a function
of cutoff $\Lambda $ for fixed values of the LAB energy, $T_{\rm
LAB}=105\: {\rm MeV}$ (upper pannel) and $T_{\rm
LAB}=300\: {\rm MeV}$ (lower pannel).  The renormalization counterterms $C_0
(\Lambda) $ and $C_2 (\Lambda) $ are always determined by fixing the
scattering length $\alpha_0 $ and the effective range $r_0 $ to their
experimental values $\alpha_0 = -23.74 {\rm fm}$ and $r_0 = 2.77 {\rm
fm}$. For $\Lambda > \Lambda_c \sim 500 {\rm MeV}$ both $C_0 (\Lambda)
$ and $C_2 (\Lambda) $ become complex while the phase shifts remain
real.}
\label{fig:1S0-ps-C0-C2-mom.300.eps}
\end{figure}

In the contact theory we found that for $\Lambda > 380 {\rm MeV} $ the
counterterms $C_0$ and $C_2$ diverge before becoming complex
numbers. Thus, the corresponding Hamiltonian becomes non-self-adjoint,
in harmony with the Wigner causality bound violations unveiled in
Ref.~\cite{Phillips:1996ae,Scaldeferri:1996nx}. When the NLO
contribution to the potential is included, the critical value of the
cutoff is slightly shifted towards the higher values $\Lambda_c=
480-550 {\rm MeV}$; above those values the short distance contribution
to the potential becomes non-self-adjoint and causality bounds are
violated. Nevertheless, and similarly to the pion-less theory, the
phase shifts remain real beyond this critical cutoff. In
Fig.~\ref{fig:1S0-ps-C0-C2-mom.eps} the value of the phase shift for
both NLO and N2LO approximations is depicted for several cutoff
values. As we see, at NLO one observes large variations when the
cutoff is changed from $\Lambda=2000 {\rm MeV}$ to $\Lambda=2780 {\rm
MeV}$. In Fig.~\ref{fig:1S0-ps-C0-C2-mom.300.eps}, we plot the value
at a fixed LAB energies as a function of the cutoff
for both NLO and N2LO. As we see the phase shift does not seem to
converge to any particular value when $\Lambda$ is increased; in fact
large variations can be clearly seen at moderate cutoff values,
$\Lambda \sim 200 {\rm MeV}$ for $E_{\rm LAB}=105 {\rm MeV}$ and
$\Lambda \sim 500 {\rm MeV}$ for $E_{\rm LAB}=300 {\rm MeV}$. On the
other hand, let us note that a plateau region not always appears, and
when it does the residual contribution from pion-exchange effects is
less important and the counterterms may dominate the calculation. In
other words, if the cutoff was too small, one would be driven back to
an effective range expansion with no visible contribution from chiral
potentials whatsoever.  Thus, any finite cutoff calculation within a
higher cutoff regime turns out to be strongly cutoff dependent, or
else the cutoff must be fine tuned to intermediate energy data, hence
becoming an esssential, and not an auxiliary, parameter of the
theory. Note that all the problems are triggered by {\it insisting} on
fitting the effective range parameter to the experimental value by
introducing the $C_2$ counterterm required by Weinberg power counting
on the short distance interaction. In contrast, if $C_2=0$, as
advocated in
Refs.~\cite{PavonValderrama:2005gu,Valderrama:2005wv,PavonValderrama:2005uj}
and Sec.~\ref{sec:N3LO} above, not only is the phase shift convergent
in the limit $\Lambda \to \infty $ (in practice $\Lambda > 1 {\rm
GeV}$) but also most of the effective range is saturated by the chiral
potential~\footnote{Actually, in the limit $\alpha_0 \to \infty$ and
under the assumption of Van ver Waals dominance one gets at N2LO the
analytical result (see Ref.~\cite{Valderrama:2005wv} for details),
$$
r_0 = \frac{16 \Gamma (5/4)^2}{3\pi} \left[ \frac{3 g_A^2}{128 \pi^2 f_\pi^4}
\left( - 4 + 15 g_A^2 + 24 c_3 M - 8 c_4 M  
\right) \right]^\frac14   
$$ which yields $r_0 \sim 2.33 {\rm fm}$, a surprissingly good
approximation. Further studies exploiting the chiral Van der Waals
correlations can be found in Ref.~\cite{Valderrama:2005wv}.}.

Although we
have checked that these results are numerically robust within the LSE
by increasing the number of grid points, there is always a reasonable
doubt, since these are demanding calculations. Thus, it would be nice
to understand the lack of convergence from a different perspective,
as we do next.

\begin{figure*}[]
\begin{center}
\epsfig{figure=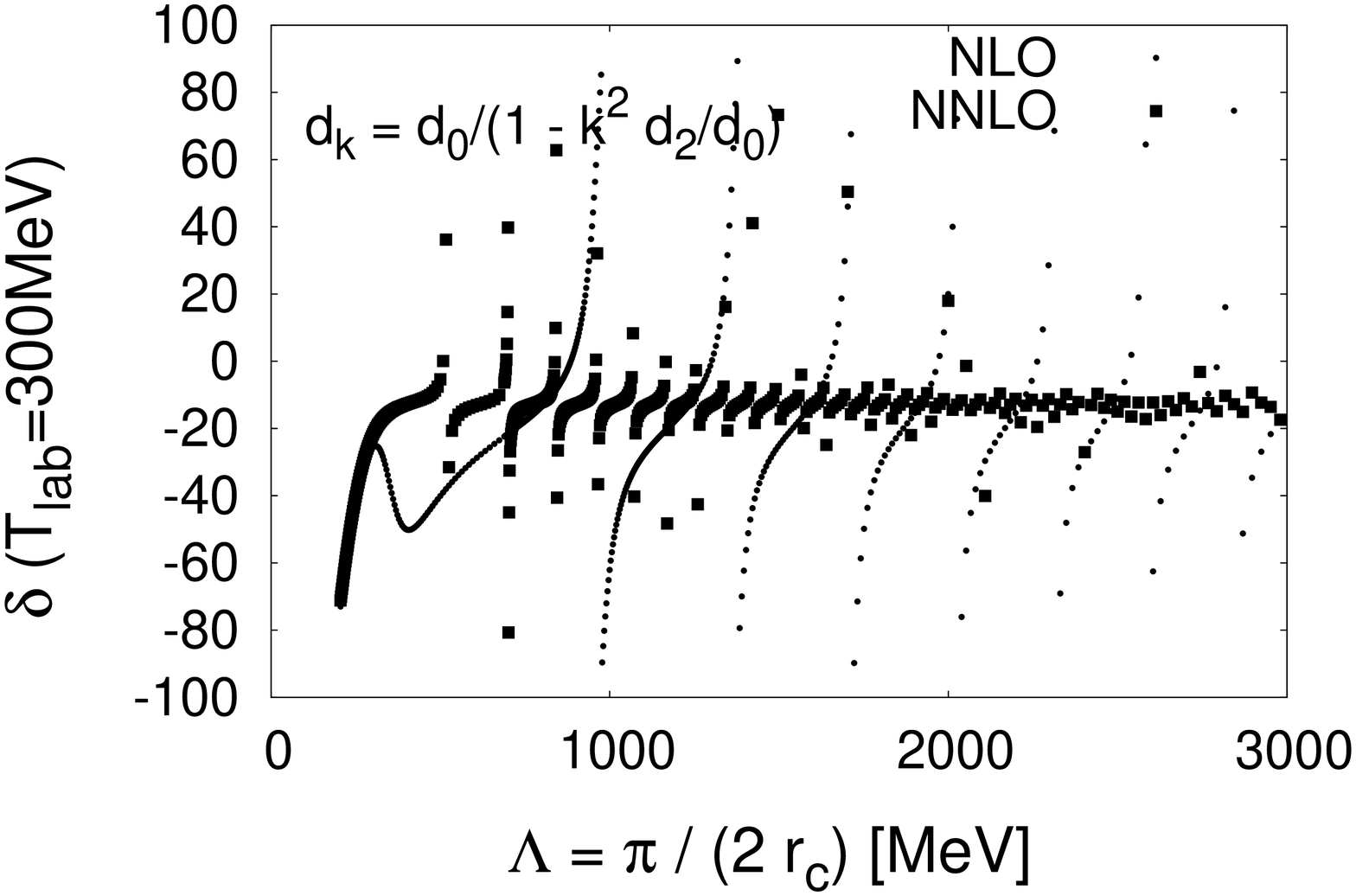,height=6.5cm,width=5.5cm}
\epsfig{figure=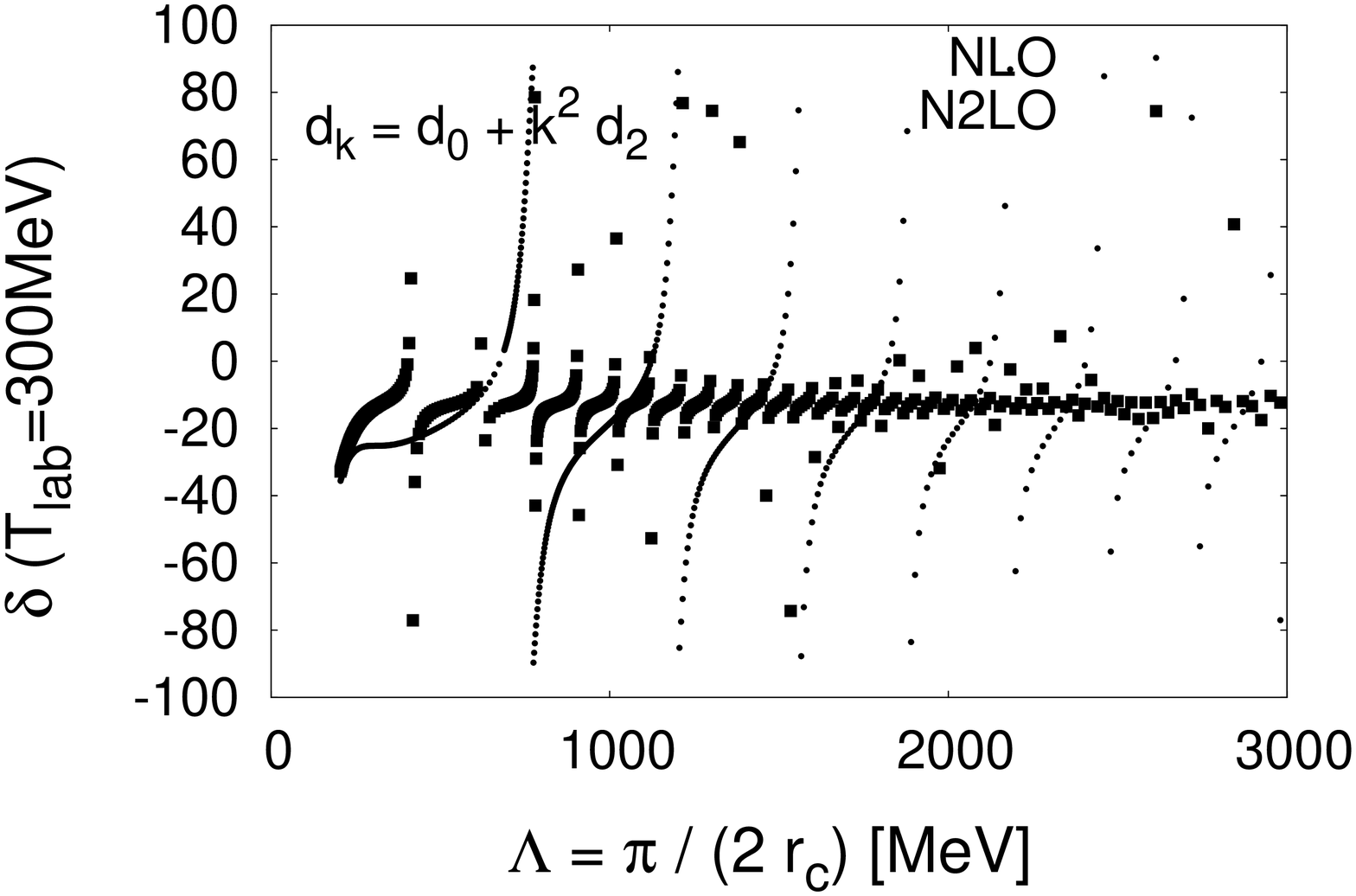,height=6.5cm,width=5.5cm}
\epsfig{figure=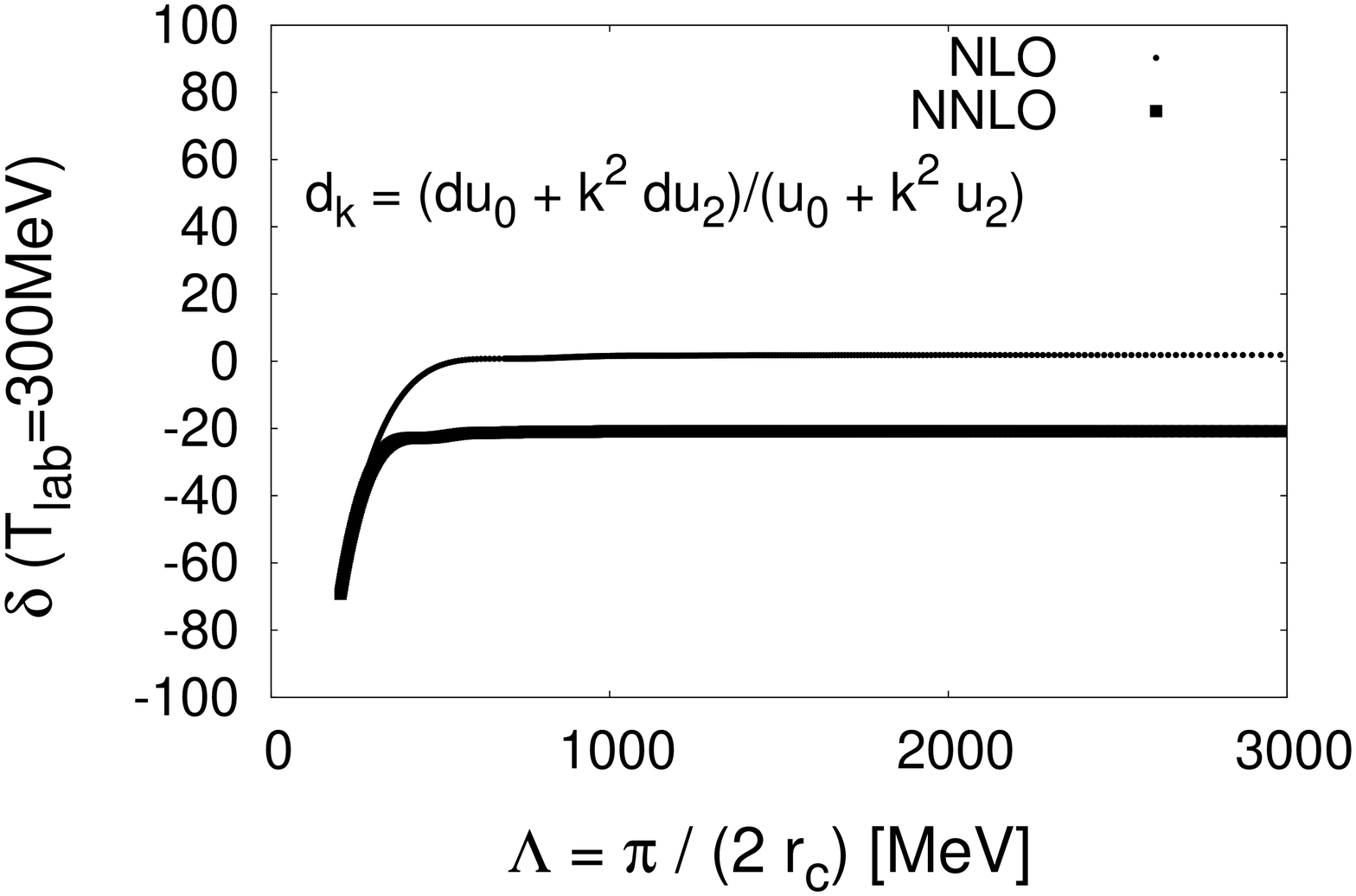,height=6.5cm,width=5.5cm}
\end{center}
\caption{NLO and N2LO running of the phase shifts as a function of the
equivalent momentum cutoff $ \Lambda = \pi / (2 r_c) $ (in MeV) with
$r_c$ the short distance cutoff, for fixed values of the LAB energy
$T_{\rm LAB}=300\: {\rm MeV}$ and for several parameterizations of the
short distance energy dependent logarithmic derivative (see
Eq.~(\ref{eq:d_p's}).  In all three cases, the chiral potential for $r
> r_c$ is the same and the scattering length $\alpha_0 $ and the
effective range $r_0 $ are both fixed to their experimental values
$\alpha_0 = -23.74\: {\rm fm}$ and $r_0 = 2.77 \: {\rm fm}$.}
\label{fig:1S0-ps-C0-C2-coor.300.eps}
\end{figure*}
\begin{figure*}[]
\begin{center}
\epsfig{figure=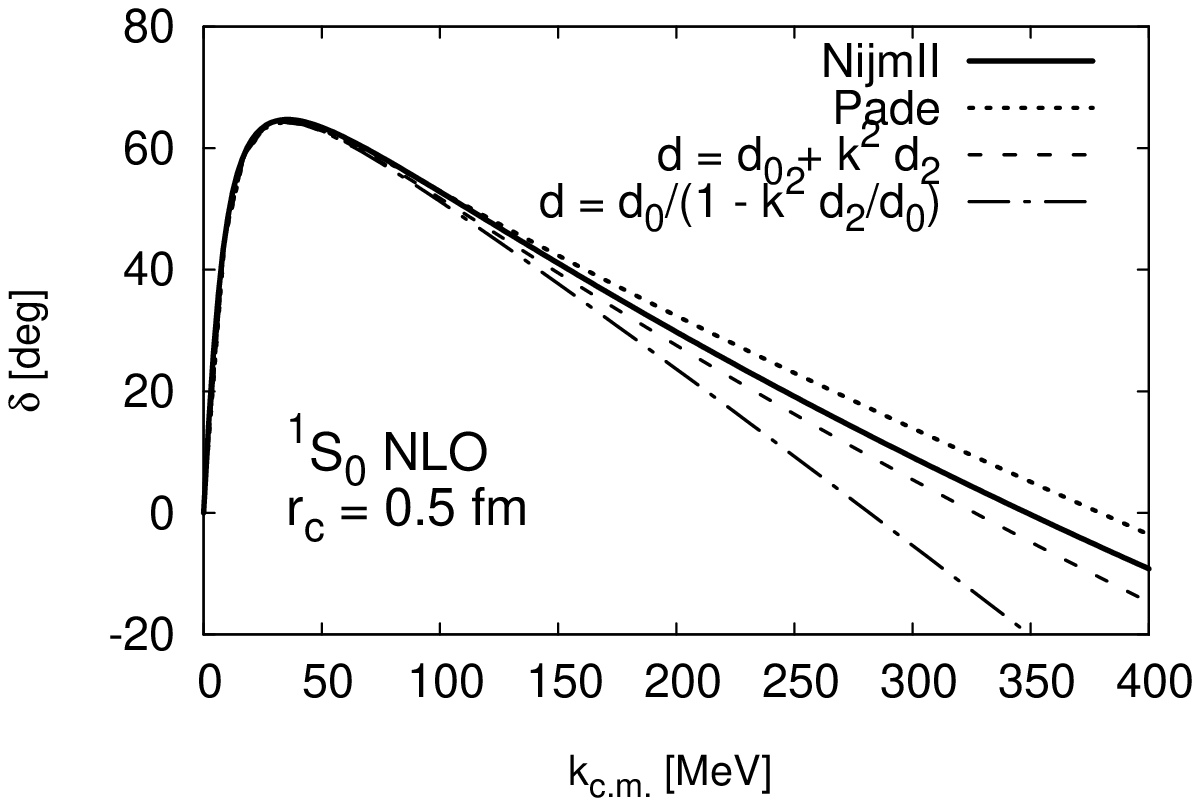,height=5.5cm,width=7.5cm}
\epsfig{figure=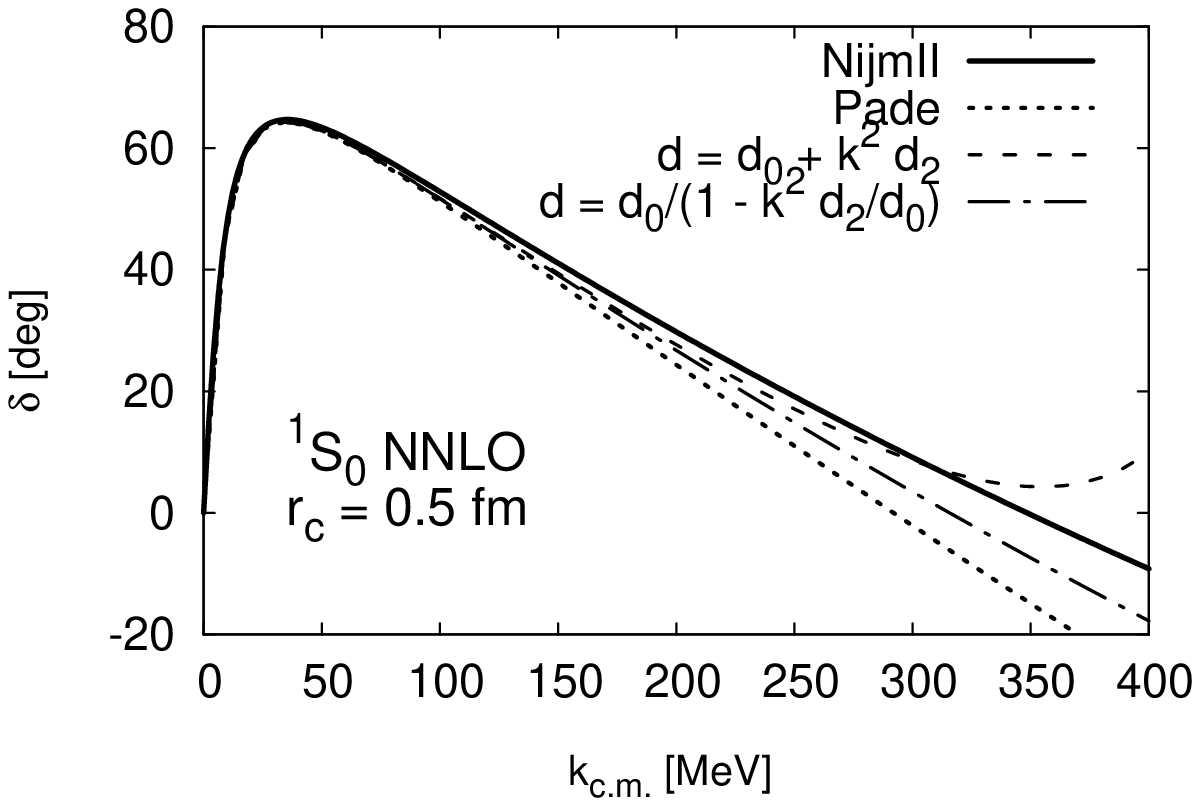,height=5.5cm,width=7.5cm}
\epsfig{figure=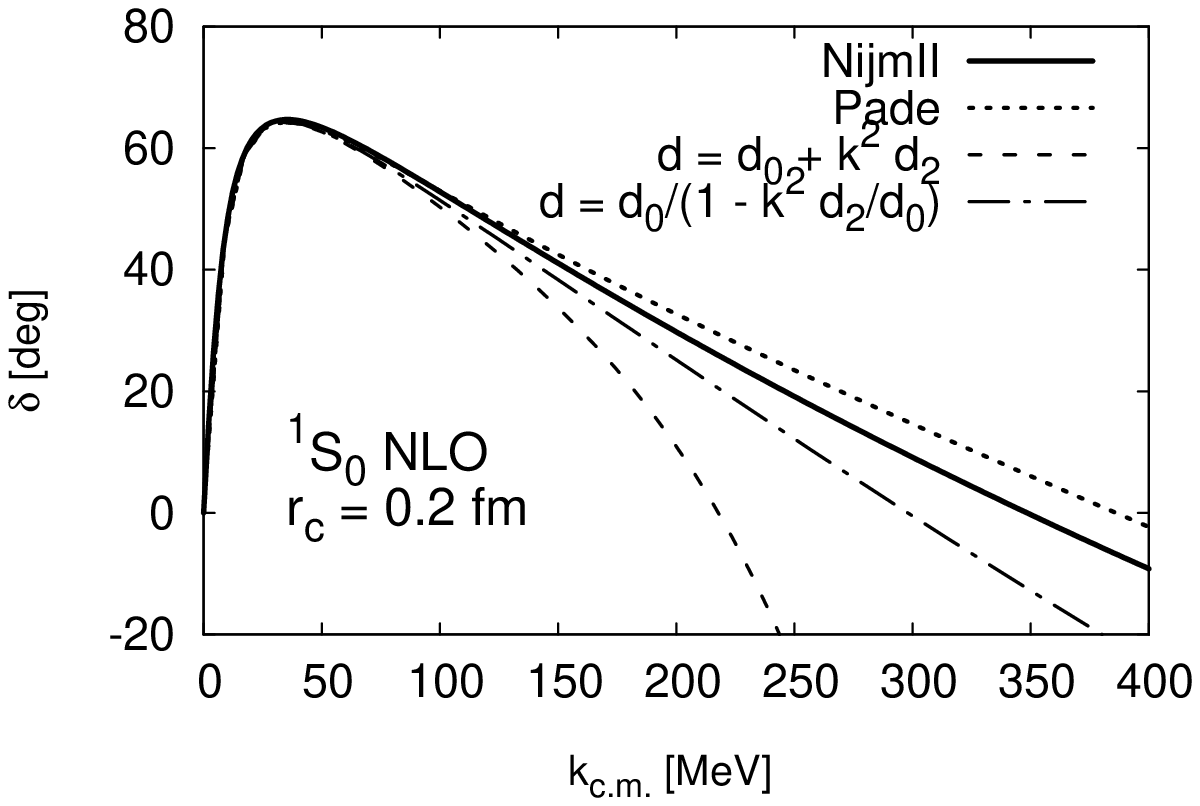,height=5.5cm,width=7.5cm}
\epsfig{figure=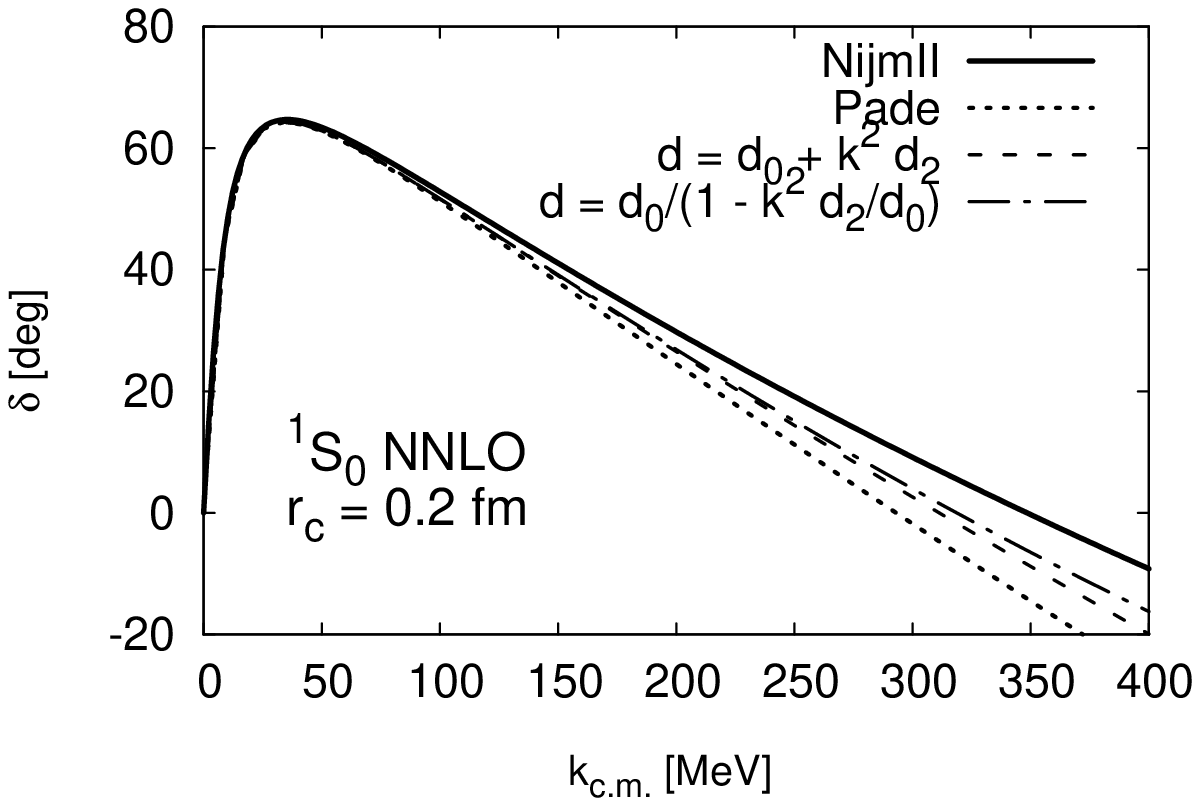,height=5.5cm,width=7.5cm}
\end{center}
\caption{NLO (left panel) and N2LO (right panel) convergence of the
phase shifts as a function of the CM momentum for fixed values of the
short distance cutoff $r_c$ for several parameterizations of the short
distance energy dependent logarithmic derivative (see
Eq.~(\ref{eq:d_p's}). In all three cases, the chiral potential for $r
> r_c$ is the same and the scattering length $\alpha_0 $ and the
effective range $r_0 $ are both fixed to their experimental values
$\alpha_0 = -23.74\: {\rm fm}$ and $r_0 = 2.77\: {\rm fm}$.}
\label{fig:1S0-ps-C0-C2-coor.eps}
\end{figure*}

\subsection{Coordinate space}

Let us compare the previous findings in momentum space with related
investigations in coordinate space~\cite{Valderrama:2007nu}. A
thorough study in coordinate space has revealed that the very existence of
the $\Lambda \to \infty $ limit may actually depend on the specific
representation of the short distance physics. This analytical result
has been verified by numerical calculations and will become
extremely helpful in analyzing the momentum space calculations
presented above. Thus, for the sake of completeness it is worth
reviewing the emerging pattern from Ref.~\cite{Valderrama:2007nu}.

As we have discussed in Sec.~\ref{sec:coor-space}, within the boundary
condition regularization the {\it unknown} short distance interaction
is represented by the logarithmic derivative of the wave function at
the short distance cutoff radius $r_c$, $ u'_k( r_c) / u_k(r_c) $. In
an energy expansion of the wave function at short distances, one has
$u_k = u_0 + k^2 u_2 + \dots $ and its logarithmic derivative for
which a continuity condition is required. This introduces an energy
dependence which eludes the Wigner causality bound discussed in
Refs.~\cite{Phillips:1996ae,Scaldeferri:1996nx}, since self-adjointness
is broken from the start. As we discussed previously in the contact
theory described in Sec.~\ref{sec:coor-pion-less}, at second order in
the energy the neglected terms are ${\cal O} (p^4)$ and {\it any} of
the three representations displayed by Eq.~(\ref{eq:d_p's}) might be
equally acceptable. Actually, we found that renormalized amplitudes
fall into two classes and that not all of them approach the
renormalized limit in the same way, see
Eq.~(\ref{eq:coor-pion-renor}). Here, we will extend that study by
inquiring what happens when the same three representations displayed
by Eq.~(\ref{eq:d_p's}) are used and the long distance potential
$V_L(r)$ is taken to be the NLO and N2LO for $r > r_c$ as the cutoff
is removed, $r_c \to 0$. Note that all three cases possess {\it by
construction} the same scattering length $\alpha_0$ and effective
range $r_0$ and the same long distance potential for $r > r_c$. Thus,
any difference is clearly attributable to the different short distance
representation. The results are displayed in
Fig.~\ref{fig:1S0-ps-C0-C2-coor.eps} for fixed short distance cutoff
values $r_c$ as a function of the CM momentum $p$. For finite values
of $r_c$ we see a difference which can naturally be explained by the
different off-shell behaviour of the short distance physics. As we see,
the difference persists as the cutoff is being removed and in fact is
magnified. As was already pointed out in a previous work~\cite{
Valderrama:2007nu}, only the case $I$ representing a rational function
turns out to yield a unique and well defined finite value for the
phase shift~\footnote{Note that in the contact theory, this
rational representation provides the softest regulator, i.e. cutoff
effects scale quadratically and not linearly as in all others, see
Eq.~(\ref{eq:d_p's}).}.

This fact is clearly seen from inspection of
Fig.~\ref{fig:1S0-ps-C0-C2-coor.300.eps} where we plot the phase
shifts for a fixed value of the CM momentum $p=300\: {\rm MeV}$ as a
function of the equivalent sharp momentum cutoff $\Lambda= \pi /( 2
r_c)$ derived for the pion-less theory in Sec.~\ref{sec:coor-mom}. The
striking similarity between the momentum space calculation presented
in Fig.~\ref{fig:1S0-ps-C0-C2-mom.300.eps} and the coordinate space
calculation displayed in Fig.~\ref{fig:1S0-ps-C0-C2-coor.300.eps} for
{\it finite cutoffs} is noteworthy although not completely surprising
in the light of the analysis of Appendix~\ref{sec:nyquist}.  There,
the finite cutoff equivalence, $\Lambda= \pi /( 2 r_c)$, deduced in
the contact theory is shown to hold also in the presence of a local
potential.

Of course, all these features depend on the singular character of the
interaction at short distances and do not depend on the specific form
of the potential, so we expect them to hold also when $\Delta$ degrees
of freedom are explicitly taken into account.

Although we cannot prove it analytically in momentum space, the
remarkable similarity of the coordinate space analysis with the
present momentum space calculations strongly suggests that the
standard polynomial representation of the short distance potential is
{\it incompatible} with renormalization. Of course, this does not
preclude the possible existence of a suitable potential representation
of the short distance interaction, most likely with complex
counterterms, such that both low energy parameters can be fixed and
the cutoff can at the same time be removed.  We leave such an
interesting investigation for the future.

\section{Conclusions and Outlook}
\label{sec:conc}

In the present paper, we have analyzed the renormalization of the
singlet np phase shift in the $^1S_0$ channel incorporating one- and two-pion
exchange effects. For the long distance potential, the standard
Weinberg scheme based on dimensional power counting is adopted to
N3LO. However, the short distance physics is parameterized in terms of
one unique energy and momentum independent counterterm whose cutoff
dependence is determined by adjusting the scattering length to its
physical value.  The present analysis is carried out in momentum space
in a somewhat complementary manner as previously done in coordinate
space~\cite{PavonValderrama:2004nb}. Actually, we have analyzed and
reproduced those results directly in the more popular momentum space
after the cutoff has been effectively removed. In order to stress
the equivalence of both approaches, we have displayed many results in
parallel. This is not just a stylistic matter of presentation; besides
numerical simplicity, much understanding of the renormalization
problem has been achieved by a direct analysis in coordinate space at
least for local potentials. The present work provides a further example
in this respect.

The momentum space framework allows a direct extension to include N3LO
contributions which include non-local pieces in the long distance
chiral potential. The main outcome of such a calculation is that the
N3LO chiral potential induces rather small corrections as compared to
the N2LO results in all of the elastic scattering region. As anticipated
previously~\cite{Valderrama:2005wv}, this happens to be so even when
the N3LO potential is more singular than the N2LO one at short
distances. Moreover, the analytical scaling behaviour for large
cutoffs $\Lambda$ of the scattering amplitude predicted in a previous
coordinate space analysis~\cite{Valderrama:2007nu} is confirmed
qualitatively by the present momentum space calculations; increasing
the order in the expansion of the long distance potential suppresses
more the finite cutoff dependence.  In summary, the scheme is
convergent, but differs from the expected phase shifts obtained from
partial wave analyses. An error analysis of the results shows that
although there are some uncertainties induced by the input parameters,
the corresponding error bands are not large enough to be compatible
with the Partial Wave Analysis of the Nijmegen group.

The N3LO approximation to the long distance chiral potential is
computed within a heavy baryon expansion and assuming only explicit
nucleon degrees of freedom in the NN potential. This implies in
particular that the $N\Delta$ splitting is considered a non-small
parameter. However, this number is about twice the pion mass, so it is
not clear whether such an assumption is fully justified. Therefore,
we have analyzed the NN scattering problem when explicit $\Delta$
intermediate state excitations are included in the potential. The net
result is that, after renormalization, the $1\Delta$ and $2\Delta$ box
diagrams contributions mimic extremely accurately the $\Delta$-less
N2LO and N3LO potentials, respectively, and thus fail to describe the
higher energy region of the $^1S_0$ phase shift. Given the fact that
these contributions also fall off at large distances as $\sim e^{- 2
m_\pi r}$, this result reinforces the conclusion that there is some
shorter range missing physics beyond that provided by TPE.

Further, it is interesting to compare the present renormalized results
with those obtained for a two pion exchange potential computed within
a relativistic baryon framework which sums up all heavy nucleon
components~\cite{Higa:2007gz} thus contains a full determination of
TPE contributions assuming all other degrees of freedom (including
explicit $\Delta$'s) are infinitely heavy. There, all np partial waves
with $j \le 5$ are analyzed and for the particular case of the $^1S_0$
channel the results are rather similar to those found here. This,
again, suppports the present conclusion that, presumably, the
components of the NN potential with a shorter range than TPE might
finally provide the missing repulsion needed to reduce the $3\%$
overshooting of the $^1S_0$ effective range as well as the too low
value of the phase shift in the region with CM momenta $p > m_\pi$.

We have also investigated the convergence of the standard Weinberg
counting for both the short distance as well as the long distance
potential. At NLO and N2LO this corresponds to including a polynomial
momentum dependence in the short distance interaction by means of two
counterterms which may then be fixed by adjusting the scattering
length and the effective range to their experimental values. We find
that, with these two renormalization conditions, the counterterms turn
out to be complex for not too large cutoffs (around $500 {\rm MeV}$),
signaling the breakdown of self-adjointness of the potential, and more
generally suggesting a violation of Wigner's causality
condition. Nonetheless, despite phase shifts remaining real, a unique
renormalized limit does not exist. This lack of convergence agrees
with similar findings in coordinate space anticipated in
Ref.~\cite{Valderrama:2007nu}. The physical understanding of the
situation is as follows.  On the one hand the long range physics is
fixed, thus, the off-shell propagation within the long range region is
unambiguous. On the other hand, the short range potential must be
adjusted to provide the threshold parameters such as the scattering
length, the effective range etc. If the short range potential is just
determined from fixing the scattering length only, this is at zero
energy and becomes zero range when the cutoff is removed, so there is
no off-shell ambiguity. In contrast, fixing further the effective
range requires non-zero energy and there are in fact infinitely many
ways of parameterizing this, with different off-shell behaviour even
when the cutoff is removed. In other words, fixing a finite range and
removing the cutoff depends on details of the method. These conflicts
between off-shellness and finiteness are not new in field theory.
Green's functions which are renormalized on-shell do not necessarily
provide finite off-shell amplitudes.  The standard Weinberg
parameterization as a polynomial in momenta is one choice which may
not turn out to be consistent with renormalizability. In this regard,
it is interesting to mention that coordinate space detailed studies
based on renormalization group properties~\cite{Valderrama:2007nu}
suggest the existence of a suitable short distance representation
yielding convergent results in theories with more than one
counterterm.  The generalization of those results to momentum space
would require an in-depth study of the renormalization group in the
presence of eventually nonlocal but singular potentials. We note that
the energy dependence of the coordinate space solution violates
self-adjointness explicitly and we expect that most likely an energy
independent non-polynomial momentum space solution would spoil
self-adjointness as well.

It is important to note that with a finite cutoff and four
counterterms the $^1S_0$ phase shift has been successfully described
within the standard Weinberg
counting~\cite{Entem:2003ft,Epelbaum:2004fk}. In this regard, it is
natural to question the usefulness of taking the infinite cutoff limit
and to carry out a renormalization process. From a physical point of
view the limit $\Lambda \to \infty $ corresponds to consider other
degrees of freedom than those not considered explicitly to be
infinitely heavy. Thus, the aim of the renormalization program is far
more stringent than previous finite cutoff calculations. Within such a
framework any failure can be unambiguously attributed to missing
physical information on the long distance potential, and the
renormalization process highlights this in a rather vivid manner as
one clearly sees when going from OPE to TPE potentials. From a
mathematical perspective the renormalizability requirement imposes
tight constraints on the admissible forms of the short distance
physics for some given long distance interactions. These are powerful
conditions which have traditionally been the great strength of the
renormalization ideas to avoid unnecessary proliferation of
interactions. Since the power counting of long distance potentials is
not uniquely determined yet (one may e.g. include or not explicit
$\Delta$'s), it would be very helpful to see what kind of short
distance constraints on those potentials might be imposed as well on
the basis of renormalizability or other principles.

The problems with the Weinberg counting for the short distance
polynomial form in the momenta of the interaction with renormalization
found in this paper for NLO and N2LO is to be added to the other ones noted in previous
works \cite{Nogga:2005hy,Valderrama:2005wv}. On the mathematical side,
it is noticeable that the coordinate space analysis of
Ref.~\cite{Valderrama:2005wv} conjectured this result by exploiting
the compelling requirement of completeness and self-adjointness for
the renormalized quantum mechanical problem for a local chiral
potential which at first sight may seem completely germane concepts to
the EFT machinery. On the phenomenological side, it should be noted
that this inconsistency result does not explain why the renormalized
phase shift with just one counterterm comes out reasonably close to
the accepted ones, but certainly makes this unique and finite
prediction more credible and inevitable from a theoretical perspective 
if self-adjointness is maintained, 
and provides further confidence on the virtues of a renormalization
principle within the chiral approach to the NN interaction.

\begin{acknowledgments}

One of us (E.R.A.) thanks Diego Pablo Ruiz Padillo (Departamento de
F{\'{\i}}sica Aplicada, Universidad de Granada) for patient
instruction on the virtues of Nyquist theorem some years ago. We also
thank Evgeny Epelpaum for a critical reading of the manuscript and
Daniel Phillips for  remarks on the paper.

The work of D. R. E. has been partially funded by the Ministerio de
Ciencia y Tecnolog\'\i a under contract No. FPA2004-05616, the Junta
de Castilla y Le\'on under contract No. SA016A07. The work of E.R.A.
is supported in part by funds provided by the Spanish DGI and FEDER
funds with grant no. FIS2005-00810, Junta de Andaluc{\'\i}a grants no.
FQM225-05, EU Integrated Infrastructure Initiative Hadron Physics
Project contract no. RII3-CT-2004-506078.  M. P. V. has been funded by
the Deutsche Forschungsgemeinschaft
(SFB/TR 16), Helmholtz Association (contract number VH-NG-222). 
R. M. has been supported in part by the U. S. National Science Foundation 
under Grant No. PHY-0099444.

\end{acknowledgments}

\appendix 

\section{Momentum vs coordinate space for finite cutoffs and local potentials}
\label{sec:nyquist}

In this appendix we discuss further the relation between momentum and
coordinate space for {\it finite cutoffs} and local potentials. 
We will show that solving the Lippmann-Schwinger equation for a 
long range potential with a sharp cutoff $\Lambda$  
is equivalent to the solution of the Schr\"odinger equation for the 
same potential in a discretized grid with $\Delta r = \pi / \Lambda$.
This equivalence reminds the previous identification between 
sharp momentum cutoff $\Lambda$ and the short distance cutoff $r_c$ 
found in the contact theory in Sec.~\ref{sec:coor-mom}, 
$\Lambda= \pi / (2 r_c) $.
We can choose the initial grid point at $r_c = \pi / (2 \Lambda)$, thus
recovering the previously mentioned equivalence.
Actually, by invoking Nyquist theorem on the cut-off LS equation, 
we deduce a discretized version of the variable phase 
equation~\cite{Calogero:1965} which allows to discuss
both the renormalization as well as the decimation 
problem of the NN-force based on chiral potentials.

\subsection{The cutoff Lippmann-Schwinger equation} 

Let us consider as a starting point the Lippmann-Schwinger equation
for $s$ wave scattering, written in the form 
\begin{eqnarray}
T (k',k) &=& V (k',k) + M\,\int_{0}^\infty dq V (k',q)
\frac{q^2}{p^2-q^2} T  (q,k) \, , \nonumber \\ 
\label{eq:lse_w}
\end{eqnarray} 
where $T(k',k)$ are the matrix elements of the $T-$ matrix between
initial and final CM momentum states $k$ and $k'$ respectively and the
corresponding potential matrix element is given by
\begin{eqnarray} 
V (k',k) = \frac2\pi \int_0^\infty dr j_0 (k' r) j_0 (k r) V(r) r^2 \, ,   
\end{eqnarray} 
for a local potential.
Since we are integrating the intermediate momentum up to infinity,
we are implicitly assuming that $V(k',k)$ is a regular potential.
The singular potential case will be discussed later.
Now, if we cutoff the potential in momentum 
space~\footnote{Cutting off the high momentum
states is not exactly the same integrating out the high energy states,
which produces a low momentum effective energy dependent ``optical''
potential. We are focusing on the long range potential here.  The
short missing distance piece could be included by a Taylor expansion
in momenta or energies (see Sec.~\ref{sec:decim}).  }, we get the
regularized potential
\begin{eqnarray} 
V_\Lambda (k',k) = \theta (\Lambda - |k'| ) \theta ( \Lambda - |k| ) V
(k',k) \, , 
\end{eqnarray} 
where $\theta (x)$ is the Heaviside step function. In coordinate space
the cutoff potential becomes 
\begin{eqnarray} 
\frac{V_\Lambda ( r ',r)}{r r'} &=& \frac{2}{\pi}
\int_{0}^\Lambda k^2 dk \int_{0}^\Lambda k'^2 dk' 
j_0 (k r) j_0 (k r') V(k,k') \, 
\nonumber \\ & \to & \frac{\delta(r-r')}{r r'}\, V(r) \, , 
\end{eqnarray} 
which is obviously nonlocal and becomes local only when $\Lambda \to
\infty$ (second line). The cutoff LSE, where $ |q| \le \Lambda $, can
be solved by standard means. In the spirit of an EFT, based on the
idea that low energy dynamics does not depend on the details at short
distances, it may perhaps be appropriate to proceed a bit
differently. Actually, if high momentum states are cutoff from the
theory the same idea should apply to small {\it resolution } scales,
i.e. regardless whether they are short or long. That means that using
{\it too much } information on the potential $V(r)$ even if it is
exactly known point-wise may be illusory at wave-lengths longer than a
given resolution, $\Delta r$,~\footnote{For example, if a potential
$V(r)$ is supplemented by highly oscillatory ripples at short
resolution scales, $ \Delta r $ the phase shifts should be insensitive
to them at $ \lambda \gg \Delta r $.} so that one can sample $V(r)$
with {\it some} resolution $\Delta r \sim 1 / \Lambda $. This obviously
reduces the number of coordinate mesh points in the integration.

\subsection{Nyquist theorem} 

In the present context, Nyquist theorem~\cite{Oppenheim} is remarkably
useful because it provides an optimal way of sampling signals which
have a bandwidth in Fourier space, i.e., functions for which
\begin{eqnarray}
F(k) = \int_{-\infty}^\infty  e^{ i k x } f(x) dx  = 0 \qquad {\rm
for} \qquad |k| \le \Lambda  \, , 
\end{eqnarray} 
then, for the original function we have
\begin{eqnarray}
f(x)=  \int_{-\Lambda}^\Lambda  e^{ -i k x } F(k) \frac{dk}{2\pi} \, .    
\end{eqnarray} 
If we define the sampling function $f_S(x)$ of the function $f(x) $
at the equidistant points $ x_n = n \Delta x $, with the optimal $
\Delta x = \pi /\Lambda $ separation,
\begin{eqnarray} 
f_S(x) = \sum_{n=-\infty}^{\infty} f( x_n ) \Delta x \delta (x-x_n)
\, ,
\label{eq:sampl}
\end{eqnarray} 
and compute its Fourier transform we get 
\begin{eqnarray} 
F_S (k) = \Delta x \sum_{n=-\infty}^{\infty} f( x_n ) e^{i k x_n} \, . 
\end{eqnarray} 
Note that for the optimal sampling $F_S (k) = F ( k) $~\footnote{As the 
original $F(k)$ is bandwidth limited to the $[-\Lambda, \Lambda]$ interval, 
it can be expressed as a Fourier sum
$$
F(k) = \sum_{n=-\infty}^{\infty}\,a_n\,e^{i k x_n}
$$
where $x_n = n \pi / \Lambda$. Thus it is trivial to see that 
$F_S(k) = F(k)$ when the sampling is done with $\Delta x = \pi / \Lambda$.
}.  
Inverting the Fourier transform we get
\begin{eqnarray}
\bar f_S(x) &=& \int_{-\Lambda}^\Lambda  e^{ -i k x } F_S(k) \frac{dk}{2\pi} \,
\nonumber \\
&=& \sum_{n=-\infty}^\infty f(x_n) \frac{\sin
\left[{\Lambda} (x-x_n) \right]}{{\pi} (x-x_n) } \Delta x \, .
\end{eqnarray} 
Thus, if we sample the function according to Eq.~(\ref{eq:sampl}), 
the following identity holds at the sampling points
\begin{eqnarray}
\bar f_S(x_n ) = f(x_n)  \, . 
\end{eqnarray} 
Hence, there is no loss of information on the sampling points if the
sampling is done equidistantly with the optimal Nyquist frequency, $
\Delta x= \pi / \Lambda $. In particular, it does not really make
sense to sample the function with smaller $ \Delta x$.  
In the next section we apply this sampling principle to the potential.

\subsection{The optimal grid for sampling the potential in coordinate space}

Nyquist theorem also applies for the special case of the Lippmann-Schwinger 
equation in s-wave scattering, although the derivation is different in some
details to the one presented in the previous section. 
For clarity, we present here these details.

We will consider first the
general case of a non-local potential, for which the Schr\"odinger equation
in s-wave reads
\begin{eqnarray} \label{eq:schroedinger_nonlocal}
-u''(r) + M\,\int_{0}^{\infty}\,dr'\,V(r, r') u(r') = k^2\,u(r) \,.
\end{eqnarray}
We can sample this non-local potential as
\begin{eqnarray}
V_S(r,r') &=& {(\Delta r)}^2\,\sum_{m, n = 0}^{\infty}
V(r_m, r_n)\,\delta(r - r_m)\,\delta(r-r_n)\,. \nonumber \\
&&
\end{eqnarray}
For the momentum space representation of the sampled potential we get
\begin{eqnarray}
V_S(k,k') &=& {(\Delta r)}^2\,\frac{2}{\pi}\sum_{m, n = 0}^{\infty}
{\hat V}_{m,n}\,j_0(k r_m)\,j_0(k' r_n)\,, \nonumber \\
&&
\end{eqnarray}
where ${\hat V}_{m,n} = r_m r_n\,V(r_m, r_n)$.
As a consequence of the cut-off $\Lambda$, the potential $V(k,k')$ 
can be expressed as a sum of spherical bessel functions
\begin{eqnarray}
V(k,k') = \sum_{m, n = 0}^{\infty}\,a_{n,m}\,j_0(k r_n)\,j_0(k' r_m)\, ,
\end{eqnarray}
where $r_n = n \pi / \Lambda$. Then by taking 
$\Delta r = \pi / \Lambda$, the sampled potential recovers the original one,
i.e. $V_S(k,k') = V(k,k')$, for $k,k' < \Lambda$. 
By Fourier-transforming $V_S(k,k')$ back to coordinate space, it can be checked
that it reproduces the original sampling points, i.e.
\begin{eqnarray}
{\bar V}_S(r_n, r_m) = {(\frac{\pi}{\Lambda} \Delta r)}^2 V(r_n, r_m) = 
V(r_n, r_m)
\end{eqnarray}
for the Nyquist sampling frequency.

In the case of a local potential, the one which interests us most, 
we sample the following way
\begin{eqnarray} \label{eq:sampl-local}
V_S(r) &=& \Delta r\,\sum_{n = 0}^{\infty}\,V(r_n)\,\delta(r-r_n)\,.
\end{eqnarray}
After double Fourier-transforming, we get
\begin{eqnarray}
{\bar V}_S(r_n, r_m) &=& V(r_n) \frac{\Lambda}{\pi} \delta_{nm}\, ,
\end{eqnarray}
which makes the potential local for the grid points. It is in fact a finite 
cut-off version of $V(r,r') = V(r)\,\delta(r-r')$ once we notice that 
$\frac{\Lambda}{\pi}\,\delta_{nm}$ $\to \delta(r - r')$.

\subsection{The Lippmann-Schwinger equation with a cut-off}

For a Lippmann Schwinger equation with a finite cut-off $\Lambda$ all 
the matrix elements become a linear combination of separable terms.
Thus the LS equation becomes a linear matrix equation which can be solved 
by standard techniques by writing
\begin{eqnarray}
T(k,k') &=& \sum_{ij} T_{ij} j_0 (k r_i ) j_0 (k r_j) \, , \\
V(k,k') &=& \sum_{ij} V_{ij} j_0 (k r_i ) j_0 (k r_j) \, , \\
\nonumber
\end{eqnarray} 
and defining propagator matrix elements  
\begin{eqnarray}
G_{ij} = \int_0^\Lambda dq \frac{M q^2}{p^2-q^2} j_0 ( q r_i ) j_0 (q r_j)\, . 
\end{eqnarray} 
Therefore we get 
\begin{eqnarray}
T_{ij} = V_{ij} + \sum_{lm} V_{il} G_{lm} T_{mj} \, .  
\label{eq:ls_grid}  
\end{eqnarray} 
This equation can be reduced to a finite-dimensional $N \times N$ linear 
algebra problem by cutting the sums to $i,j = N$.
The effect of this simplification can be seen by sampling $V_{ij}$ in 
coordinate space (see Eqs.~(\ref{eq:sampl}) and (\ref{eq:sampl-local}))
\begin{eqnarray} 
V(r) &\to& \Delta r \sum_{n=0}^N V(r_n) \delta(r-r_n) \,\, , 
r_n = \frac{n\pi}{\Lambda}  \, , 
\end{eqnarray}
from which we can check that cutting the sum is equivalent to introducing
the (harmless) infrared cut-off $r_N$.
Note that for local potentials, the sampled potential matrix elements 
at the grid points are diagonal
\begin{eqnarray}\label{eq:V_ij_nyquist}
V_{ij} = \frac{2}{\pi}\,r_i^2\,V(r_i)\Delta r\,\delta_{ij} = 
V_i\,\delta_{ij}  \, . 
\end{eqnarray} 
Thus Eq.~(\ref{eq:ls_grid}) becomes
\begin{eqnarray}
T_{ij} = V_i \delta_{ij} + \sum_{m} V_i G_{im} T_{mj} \, ,    
\end{eqnarray} 
which looks like a multiple scattering equation, with on-shell
propagation between delta-shell scatterers.  After matrix inversion,
the on-shell solution is then given by
\begin{eqnarray}
T(p) = \sum_{ij} T_{ij} (p) j_0 (p r_i ) j_0 (p r_j)  \, . 
\end{eqnarray}

\subsection{The discretized Schr\"odinger equation}

In the previous section we have seen that for the Lippmann-Schwinger 
equation with a cut-off $\Lambda$, we can either use the original momentum 
space potential $V(k,k')$ and solve by standard means, 
or expand this potential, i.e. use the sampled potential $V_S(k,k')$,
and solve as a linear algebra problem (as they are both the same potential 
for momenta below the cut-off).

Alternatively we can directly solve the Schr\"odinger equation for the sampling
potential $V_S(r)$. This procedure will give an excellent approximation to the
solution of the LS equation (although not the exact solution, as explained at 
the end of this section) but at a much smaller computational cost. 
For this purpose we make the replacement
\begin{eqnarray}\label{eq:pot-delta-shell}
V(r) \to V_S (r)= \sum_{i=0}^N \Delta r \delta (r-r_i) V (r_i) \, , 
\end{eqnarray}
which is a superposition of equally spaced delta-shell potentials.
It should be noted that the point $r_0 = 0$ does not contribute, 
as it lies at the integration boundary of the Schr\"odinger equation. 
Then, we have
\begin{eqnarray}
\Delta r = \frac{\pi}{\Lambda} \qquad \, \qquad r_N \simeq N \Delta r \, .
\end{eqnarray} 
Thus, for a potential of size $a$ where we do not want to describe
energies higher than $\Lambda$ we should do with an infrared cut-off larger
than the potential's length scale, $r_N \gg a$, or equivalently
$ N \gg \Lambda a / \pi $. We can solve the Schr\"odinger equation piecewise, 
\begin{eqnarray}
u(r) &=& A_i \sin ( k r + \delta_{i-1/2} ) \, , \\  & & r_{i-1} \le r \le
r_i \, , \nonumber 
\end{eqnarray} 
where $A_i$ is the amplitude and $\delta_{i-1/2} $ can be understood
as the accumulated phase-shift due to adding a new delta shell at $
r_i $ chosen to be located at the midpoint $ r_{i + \frac12}$ (for reasons
to become clear soon). Note that with this choice the lowest possible
location of the phase-shift, $\delta_{1/2}$ corresponds to the point
\begin{eqnarray} 
r_c = \frac{\Delta r}2 = \frac{\pi}{2 \Lambda} \, , 
\label{eq:R_s-lambda}
\end{eqnarray} 
which we may identify with a short distance (ultraviolet)
cutoff. This identification between the momentum-space cutoff agrees
with the one obtained for the pure short range theory by
solving the LSE without any discretization (see
Sec.~\ref{sec:coor-mom}).

It should be noted that the previous method for solving the Schr\"odinger
equation does not really generate the exact solution of the LS equation 
with a cut-off, but only a close approximation. 
Although $V_S(r)$ is the best sampling function for $V_{\Lambda}(k,k')$, 
it is not the same quantum mechanical potential.
The solution to the LS equation with $V_{\Lambda}(k,k')$ can
be exactly reproduced by solving the non-local Schr\"odinger equation,
Eq.~(\ref{eq:schroedinger_nonlocal}), with the potential ${\bar V}_S(r,r')$
which comes from inverse Fourier transforming $V_S(k,k')$. The Schr\"odinger
equation should be solved with trivial initial conditions, $u(0) = 0$,
as all the physically relevant information is included in ${\bar V}_S(r,r')$.

On the contrary, if we solve $V_S(r)$, a sum of delta shells, we must include a
non trivial initial condition~\footnote{Specially since we are effectively 
ignoring the $r_0 = 0$ sampling point.}, 
even in the absence of any short range physics.
If we solve the discretized Schr\"odinger equation with a trivial initial 
boundary condition, $\delta_{1/2} = 0$, then a certain error will be 
included in the final solution. 
In the worst case we can expect an error of order $\Delta r$.
By means of the variable phase equation~\cite{Calogero:1965} we can perform a 
better assessment of the error~\footnote{For small enough $\Delta r$, the
variable phase equation will yield 
$$\delta_{\frac{1}{2}} \simeq - k\,\int_{0}^{\frac{\Delta r}{2}}\,M\,V(R)\,
dR\,.$$
As a curiosity, we can see that $\delta_{1/2}$ will scale as an inverse power 
of $\Delta r$ for a singular potential, thus signaling the need of a 
counterterm.
}, 
yielding, for example, an 
${\mathcal O}((\Delta r)^2)$ error for a Yukawa potential or 
${\mathcal O}((\Delta r)^3)$ for a square well.

From the previous discussion, it is apparent how to include a counterterm
in the computation. It enters throught the initial sampling point of $V_S$, 
which maps onto a constant term in $V_S(k,k')$~\footnote{For example, the 
$C_0$ counterterm when projected to the s-wave takes the form
$$
C_0\,\frac{\delta(r)}{4\pi r^2} \to C_0 \frac{1}{4\pi\,\Delta r\,r_0^2} \quad 
\mbox{(when discretizing).}
$$
Thus, if we write the sampled potential $V_S(k,k') = \sum_{ij}\,V_{ij}\,
j_0(k r_i) j_0(k r_j)$, and take into account Eq.~(\ref{eq:V_ij_nyquist}),
we see that $C_0$ maps only onto $V_{00} = \frac{C_0}{2 \pi^2}$, giving a 
constant contribution in momentum space, as expected.
}. 
Thus it is ignored when solving the Schr\"odinger equation for $V_S$, 
and it enters through the initial condition $\delta_{1/2}$. 
Similar remarks can be made for singular potentials, in which a counterterm 
must be included in order to obtain a stable result for $\Delta r \to 0$.

\subsection{The discrete variable phase equations} 

The previous discussion can be elaborated further to reach
interesting results.  Matching the wave functions at the points where
the delta shells are located, $r=r_i$, we simply get
\begin{eqnarray}
k \cot ( k r_i + \delta_{i+1/2} ) - k \cot ( k r_i + \delta_{i-1/2} )
&=& \Delta r U(r_i) \, . \nonumber \\
\label{eq:vf_dis} 
\end{eqnarray} 
where $U(r) = 2 \mu V(r) = M V(r)$ is the reduced potential.  This is
a recursion relation for the phase-shift at the interval midpoint $
r_{i+\frac12} = (i+\frac12) \pi / \Lambda$. Unlike the matrix equation
which has traditional storage limitations for large number of grid
points $N$, this equation does not posses this shortcoming allowing
for rather large $N$ values. Similar equations were deduced many years
ago~\cite{PhysRevC.6.1467} as a practical tool to attack the inverse
scattering problem and to determine the NN potential on the grid
points.

Defining the discretized effective range function,
\begin{eqnarray}
M_i = k \cot \delta_i \, , 
\label{eq:M_i}  
\end{eqnarray} 
we get 
\begin{eqnarray}
\frac{M_{i+\frac12} k \cot kr_i - k^2}{M_{i+\frac12}+ k \cot k r_i } &-& \frac{
M_{i-\frac12} k \cot kr_i - k^2}{M_{i-\frac12}+ k \cot k r_i } = \Delta r U
(r_i) \nonumber \\
\label{eq:m_dis}
\end{eqnarray}
which can be rewritten as a continuous fraction. 
Note that for the cutoff theory both Eqs.~(\ref{eq:vf_dis}) and
(\ref{eq:m_dis}) are {almost exact}. The only approximation comes from
the finiteness of the cutoff $\Lambda$. An important property which
will be used later on is the reflection property, namely the symmetry
under the replacement
\begin{eqnarray} 
\Delta r \to - \Delta r \qquad , \qquad \delta_{i+\frac12} \to
\delta_{i-\frac12}
\end{eqnarray} 
which means that on the grid running the relation upwards or downwards
are inverse operations of each other. Obviously this property may fail
in practice only due to accumulation of computer round-off errors over
large evolution distances. This is also the reason why we choose the
midpoint $r_{i+1/2}$ for the accumulated phase shift: in any other case
we would lose the reflection property.

The regular solution at the origin reads 
\begin{eqnarray}
\delta_{\frac12} (k)=0 \qquad \delta_{N+\frac12} (k)= \delta(k)
\end{eqnarray} 
If we take the limit $\Lambda \to \infty $ we can define $\delta ( k ,
r_i ) = \delta_i (k) $, to get
\begin{equation}
\frac{ d \delta (k,R) }{dR} = -\frac1k U(R) \sin^2 (k R+ \delta(k,R))
+ {\cal O} ( \Delta r^2 )
\, . 
\label{eq:vf}
\end{equation}
which is the variable phase equation~\cite{Calogero:1965} up to finite
grid corrections and can be interpreted as the change in the
accumulated phase when a truncated potential of the parametric form $U
(r) \theta (R-r)$ is steadily switched on as a function of the
variable $R$. Eq.~(\ref{eq:vf_dis}) is thus a discretized variable
phase equation, corresponding to a discretized potential sampled at
the optimal Nyquist frequency. This equation and its generalization to
coupled channels has extensively been used to treat the
renormalization problem in NN scattering in
Refs.~\cite{PavonValderrama:2003np, PavonValderrama:2004nb}. 

The discrete equations for the low energy parameters can be easily
computed from the low energy expansion
\begin{eqnarray} 
k \cot \delta_i (k) = -\frac1{\alpha_{0,i} } + \frac12 r_{0,i} k^2 +
v_{2,i} k^4 \cdots
\end{eqnarray} 
which yields
\begin{widetext} 
\begin{eqnarray} 
\frac1{r_i - \alpha_{0,i+\frac12}} - \frac1{r_i - \alpha_{0,i-\frac12}} &=& \Delta
r\, U(r_i) \label{eq:alpha_dis} \\ \frac{6\,\alpha_{0,i+\frac12}\,r_i^2 -
2\,r_i^3 + \alpha_{0,i+\frac12}^2\,\left( -6\,r_i + 3 r_{0,i+\frac12} \right) }
{6\,{\left( \alpha_{0,i+\frac12} - r_i \right) }^2} &=&
\frac{6\,\alpha_{0,i-\frac12}\,r_i^2 - 2\,r_i^3 + \alpha_{0,i-\frac12}^2\,\left(
-6\,r_i + 3 r_{0,i-\frac12} \right) }{6\,{\left( \alpha_{0,i-\frac12} - r_i
\right) }^2 } \label{eq:r_dis}
\end{eqnarray} 
\end{widetext} 
and similar equations for higher low energy parameters, $v_2,v_3,v_4,
\dots$.  Note that the potential only enters explicitly in the
equation involving the discretized scattering length.  One of the nice
features of this equation is the way how it handles the case of
singular points, when $\alpha_0(R)$ or other parameters diverge, since
standard integration methods for the corresponding differential
equation are based on the smoothness of the solution and hence
fail. 

The way to proceed in practice is quite simple.  For a given
number of grid points $N$ we take the scattering length,
$\alpha_{0,N+1/2} = \alpha_0 $, the effective range, $r_{0,N+1/2}=r_0 $ and
run Eq.~(\ref{eq:alpha_dis}) and Eq.~(\ref{eq:r_dis}) to determine
$\alpha_{0,1/2}= \alpha_0 (\pi /2 \Lambda) $ and $ r_{0,1/2}= r_0(\pi /2
\Lambda)$. Then, we use, e.g. $M_{1/2}= -1/\alpha_{0,1/2}+ r_{0,1/2} k^2
/2 $ and run Eq.~(\ref{eq:m_dis}) upwards to obtain $ M=k \cot \delta
= M_{N+1/2} $. Due to the reversibility of the algorithm one {\it
exactly} has $ M \to -1/\alpha_0+ r_0 k^2 /2 $ in the limit $k \to 0$
for any finite grid size $\Delta r $ (modulo computer arithmetic
round-off errors). In this way we can fix the initial boundary 
conditions to exactly reproduce any given scattering length, 
effective range, etc.

This method was used successfully~\cite{Valderrama:2005ku} to extract 
low energy threshold parameters in all partial waves with $j \le 5 $ 
from high quality potentials~\cite{Stoks:1994wp}. On that case, 
we ran Eq.~(\ref{eq:alpha_dis}) and Eq.~(\ref{eq:r_dis}) with
trivial initial boundary conditions $\alpha_{0,1/2} = 0$,
$r_{0,1/2} = 0$, etc, and obtain the potential's threshold 
parameters as $\alpha_0 = \alpha_{0, N + 1/2}$, $r_0 = r_{0, N + 1/2}$, etc.


As we have said the discretized running parameters located at the
lowest possible radius $r_c = \pi /2 \Lambda $ correspond to the short
distance interactions. Nowhere in the equations does the potential at
the origin appears explicitly, since according to
Eq.~(\ref{eq:m_dis}), one starts with $ U(\pi /\Lambda) $. From this
point of view the treatment of regular and singular potentials at the
origin is on equal footing. 

\subsection{The decimation process in momentum space} 
\label{sec:decim}

The previous equations provide the accumulated phase shifts due to the
addition of equidistant delta-shells sampling the original potential
in coordinate space. We want to show that they actually provide a
solution of the decimation problem of the LSE where the low energy
states are cutoff.  Assuming that we have the LSE with a given
cutoff $\Lambda$ related to potential, we ask how does the physical
phase-shift change when we make the transformation $\Lambda \to
\Lambda /2 $, in the potential fixed. Applying the process explained
above based on Nyquist theorem we see that this corresponds to double
the grid resolution $ \Delta r \to 2 \Delta r $. Obviously, by
repeating the process we may effectively have $ \Delta r \gg a $
(being $a$ the range of the potential) and
hence the potential never contributes and $\delta_{N+1/2} =
\delta_{1/2} $.  Thus, for $ \Lambda a / \pi \ll 1 $, the short range
theory is recovered. In the opposite limit $ \Lambda a / \pi \gg 1 $
we get the full short range plus long range physics.

If we now set a grid with a fixed number of points $ N \Delta r = N
\pi / \Lambda = a $, then making $ \Lambda \to \Lambda /2 $
corresponds to $a \to 2 a $ so it could be viewed as switching on the
potential. 

The discretized representation of the LSE is algebraically
closed, since it corresponds to a separable interaction, but any term
contains {\it all} powers in momentum. Moreover, they also vanish for
large $k$, unlike any polynomial approximation to it. The numerical
procedure explained above of computing the phase-shift from some low
energy parameters like $\alpha$,$r_0$, etc. and the long distance
potential $V(r)$ can be also carried out by assuming that the
potential at the lowest grid point, $ r_c = \pi / 2 \Lambda$, is
energy dependent yielding
\begin{eqnarray}
V_S (k,k') = V(p) j_0 \left( \frac{\pi k }{2 \Lambda} \right) j_0 \left(
\frac{\pi k' }{2 \Lambda} \right)
\end{eqnarray} 
Expanding in powers of momentum we can identify the terms 
\begin{eqnarray}
V_S (k,k') = V(p) \left[ 1- \frac{\pi^2}{12 \Lambda^2} \left( k^2 +
{k'}^2 \right)+ \dots \right]
\end{eqnarray}
with $C_0$, $C_2$, etc. 
Note that the radius of convergence of the expansion in $k$ and $k' $
is the whole complex plane, due to the meromorphic character of the
spherical Bessel functions.  


\end{document}